\documentclass[%
 reprint,
 amsmath,amssymb,
 aip,
 cha,
]{revtex4-2}

\usepackage[utf8]{inputenc}
\usepackage[T1]{fontenc}
\usepackage{graphicx}
\usepackage{dcolumn}
\usepackage{bm}
\usepackage[hidelinks]{hyperref}

\usepackage{color}
\usepackage{dsfont}

\newcommand{\heading}[1]{\medskip\noindent\emph{#1}}

\begin{document}

\title{Finite-state Markovian surrogates for long-time neuronal state distributions and firing rates}

\author{Zhongyi Wang}
\affiliation{Courant Institute of Mathematical Sciences, New York University, New York, New York 10003, USA}

\author{Louis Tao}
\email{taolt@mail.cbi.pku.edu.cn}
\affiliation{Center for Bioinformatics, National Laboratory of Protein Engineering and Plant Genetic Engineering, School of Life Sciences, Peking University, Beijing 100871, China}
\affiliation{Center for Quantitative Biology, Academy for Advanced Interdisciplinary Studies, Peking University, Beijing 100871, China}

\author{Zhuo-Cheng Xiao}
\email{xiao.zc@nyu.edu}
\affiliation{NYU-ECNU Institute of Brain and Cognitive Science, New York University Shanghai, Shanghai 200124, China}
\affiliation{NYU-ECNU Institute of Mathematical Sciences, New York University Shanghai, Shanghai 200124, China}

\date{August 24, 2026}

\begin{abstract}
Spiking neuronal networks connect cellular and synaptic mechanisms to collective activity, but estimating their long-time statistics often requires costly spike-by-spike simulation. We construct finite-state Markovian surrogates for neuronal populations with one- to three-dimensional intrinsic dynamics. A finite phase-space partition represents each population by its occupancy across single-neuron states, and the firing rate is defined uniformly as probability flux through model-specific spike transitions. We compare two closures: Type I enforces a stationary mean synaptic drive, whereas Type II retains its temporal evolution. In the leaky linear integrate-and-fire (LIF) benchmarks, the Type I estimator is accurate in temporally homogeneous cases but misses rate changes driven by synaptic timescales. In the synaptic-timescale sweep, the Type II estimator achieves small relative errors in most cases. Exponential integrate-and-fire (EIF) results exhibit an analogous qualitative pattern. FitzHugh--Nagumo (FHN) and reduced Hodgkin--Huxley (HH) examples extend the construction from one-dimensional threshold--reset dynamics to continuous trajectories in nonlinear two- and three-dimensional phase spaces. Together, these examples show how the finite-state construction generates model-dependent estimates of long-time state distributions and firing rates without resolving every neuron and spike.
\end{abstract}

\maketitle

\begin{quotation}
Large neuronal networks can occupy distinct long-time activity regimes, but finding them by repeatedly simulating every spike is costly. We replace the full network with a reduced description comprising a finite-state distribution over neuronal phase space and, for recurrent spike-driven models, a dynamical equation for the mean synaptic input. Our experiments cover spiking neuronal networks with single-neuron models ranging from one to three dimensions. We show that retaining the temporal evolution of the mean synaptic input substantially improves firing-rate estimates when the stationary closure is insensitive to synaptic timescales.
\end{quotation}

\section{Introduction}
\label{sec:intro}

Spiking neuronal networks provide a mechanistic link from cellular and synaptic properties to collective neural dynamics \cite{wang2002probabilistic,izhikevich2008large,eliasmith2012large,potjans2014cell,markram2015reconstruction,bezaire2016interneuronal,chariker2016orientation,schmidt2018multi,billeh2020systematic,siegle2021survey,zenke2021remarkable,chariker2022computational}. The same network can move between asynchronous, rhythmic, and high-activity states as its physiological parameters vary. Mapping these parameters to network behavior normally requires repeated simulation of many state variables and spike events, which becomes expensive as network size and parameter dimension increase \cite{schmitt2023simulation,brian2genn,kulkani2021benchmark}. Analytical reduction is difficult for a complementary reason: spikes act discontinuously on postsynaptic neurons and, in threshold--reset models, irreversibly change the state of the presynaptic neuron. Thus nonlinear single-neuron dynamics and event-driven network interactions are difficult to separate in a faithful reduction.

In many applications, however, the target is not an individual spike sequence but a long-time statistic. Firing rates and invariant state distributions summarize persistent activity and provide natural quantities for comparing model predictions with experimental observations \cite{mizuseki2013,kass2023,dayanabbott,wilson1972excitatory}. Neural-field, mean-field, kinetic, and Fokker--Planck descriptions have made such statistics substantially more tractable \cite{brunel2000dynamics,brunel1999fast,wilson1972excitatory,NykampTranchina2000,cai2006kinetic,MontbrioEtAl2015,baccelli2021pairreplica}. Many such reductions become most effective under weak coupling, large-network limits, or homogeneous inputs. Here we ask how much temporal feedback between population activity and synaptic input must be retained in finite networks to distinguish a change in average drive from a change in its temporal organization.

A second challenge is to extend the same reduction beyond one-dimensional threshold--reset neurons. For leaky linear and exponential integrate-and-fire (LIF and EIF) models, the population state can be represented by a voltage distribution supplemented by a refractory state. FitzHugh--Nagumo (FHN) and reduced Hodgkin--Huxley (HH) neurons instead evolve along continuous trajectories in two- or three-dimensional phase spaces \cite{fitzhugh1961impulses,hodgkin1952quantitative}. Their long-time distributions occupy nonlinear subsets of these spaces, and a spike is detected along a trajectory rather than defined by an imposed reset. This changes both the state to be approximated and the definition of firing.

We address both challenges by extending an earlier Markovian reduction for LIF networks \cite{cai2021model,wu2023multi,chang2025minimizing}. In the spirit of Ulam's approximation, we partition the single-neuron phase space to obtain a finite-state representation of population occupancy. A model-dependent generator combines intrinsic dynamics with external and recurrent input transitions, while model-specific firing transitions are tagged for flux calculation. Firing rate is then the probability flux through tagged firing transitions. The construction spans a hierarchy of intrinsic complexity, from one-dimensional reset models to multidimensional continuous spike-generating models. A second hierarchy controls how much temporal information is retained. The Type I estimator solves a stationary self-consistency problem for the population distribution and the mean synaptic drive. The Type II estimator evolves their coupled dynamics before taking a long-time average.

The LIF network provides the main quantitative benchmark. In the Type I estimator, the synaptic timescale cancels from the mean event rate, so the stationary closure cannot reproduce rate changes caused by temporal synaptic filtering. The Type II estimator restores this dependence by evolving the mean pending-input pools. EIF model results exhibit an analogous qualitative pattern. The multidimensional examples then test whether the same occupancy-and-flux construction remains usable beyond imposed threshold--reset dynamics. Together, the results extend the earlier LIF estimator across synaptic closures and intrinsic neuron models.

\section{Neuronal network models}
\label{sec:models}

\subsection{Common architecture and input dynamics}

We consider finite networks containing $N_E$ excitatory neurons and $N_I$ inhibitory neurons; the total size is $N=N_E+N_I$. Neurons are homogeneous within each population. The state of neuron $i$ in population $X\in\{E,I\}$ is
\begin{equation}
 \bm z_i^X(t)=\left(V_i^X(t),u_{i,2}^X(t),\ldots,u_{i,d_X}^X(t)\right)
 \in\Omega_X\subset\mathbb R^{d_X},
 \label{eq:state}
\end{equation}
where $V_i^X$ is a membrane potential or voltage-like variable and the remaining components are recovery or gating variables. Between discrete input or reset events, the state obeys
\begin{equation}
 \frac{d\bm z_i^X}{dt}=\bm F_X(\bm z_i^X)+\bm b_X J_i^X(t,V_i^X),
 \label{eq:common_dynamics}
\end{equation}
where $\bm F_X$ is the intrinsic vector field. Inputs act in the voltage direction: $\bm b_X=\bm e_V=(1,0,\ldots,0)^{\mathsf T}$ for the unit-capacitance forms used for LIF, EIF, and FHN, and $\bm b_X=\bm e_V/C^X$ for the HH equations. Specific forms of $\bm F$ are given in Appendix~\ref{app:model_details}.

The combined recurrent and external drive is decomposed as
\begin{equation}
 J_i^X(t,V)=J_{i,\mathrm{rec}}^X(t,V)
 +I_i^{X,\mathrm{ext}}(t),
 \label{eq:input_current}
\end{equation}
where the form of the recurrent contribution depends on the neuronal model. The superscript \(XY\) denotes input or recurrent coupling from a neuron in population \(Y\) to a neuron in population \(X\).

\heading{External input.} Each neuron is driven by an independent Poisson train of rate $\lambda^{X,\mathrm{ext}}$ and a constant current $I_i^0$,
\begin{equation}
 I_i^{X,\mathrm{ext}}(t)=S^{X,\mathrm{ext}}\sum_\mu
 \delta(t-t_{i,\mu}^{X,\mathrm{ext}})+I_i^0.
 \label{eq:external_input}
\end{equation}
Under the within-population homogeneity assumed here, $I_i^0\equiv I_0^X$. We set $I_0^X=0$ in the LIF and EIF tests. 

\heading{Spike transmission.} LIF, EIF, and HH models use chemical synapses to transfer spikes, and the recurrent current is conductance-based:
\begin{equation}
 J_{i,\mathrm{rec}}^X(t,V)
 =g_i^{XE}(t)(V^E-V)+g_i^{XI}(t)(V^I-V),
 \label{eq:conductance_current}
\end{equation}
where \(V^E\) and \(V^I\) are the excitatory and inhibitory reversal potentials. Under the spike-by-spike delivery convention considered here,
\begin{equation}
 g_i^{XY}(t)=S^{XY}\sum_{\substack{j\in Y\\j\ne i}}
 \sum_{\mu}A_{ij\mu}^{XY}G^{XY}(t-t_{j,\mu}^Y),
 \label{eq:synaptic_conductance}
\end{equation}
where \(A_{ij\mu}^{XY}\) is a Bernoulli delivery indicator with probability \(P^{XY}\), \(S^{XY}\) is the delivered strength, and \(t_{j,\mu}^Y\) is the \(\mu\)th spike time of neuron \(j\). The synaptic kernel is
\begin{equation}
 G^{XY}(t)=\frac{1}{\tau^{XY}}
 \exp\left(-\frac{t}{\tau^{XY}}\right)\mathcal H(t),
 \label{eq:synaptic_kernel}
\end{equation}
with synaptic timescale \(\tau^{XY}\) and Heaviside function \(\mathcal H\).

\heading{Gap-junction coupling.} We test electrical coupling through gap junctions in the FHN model. The recurrent interaction is
\begin{equation}
 J_{i,\mathrm{rec}}^X(t,v_i^X)
 =
 \sum_{Y\in\{E,I\}}\sum_{j\in Y,j\neq i}
 G_{\mathrm{gap}}^{XY}
 \left[v_j^Y(t)-v_i^X(t)\right].
 \label{eq:fhn_gap_current}
\end{equation}
Here, \(G_{\mathrm{gap}}^{XY}\) is the time-independent electrical-coupling strength from population \(Y\) to population \(X\). Unlike the recurrent input in Eq.~\eqref{eq:conductance_current}, the FHN coupling acts continuously through the instantaneous voltage differences and does not involve reversal potentials, spike-triggered conductance increments, or synaptic decay kernels. Consequently, upward threshold crossings are used only to measure the FHN firing rate; they do not trigger recurrent coupling through spikes.

The voltage component of Eq.~\eqref{eq:common_dynamics} becomes
\begin{equation}
 \frac{dv_i^X}{dt}
 =
 F_{X,v}(\bm z_i^X)
 +J_{i,\mathrm{rec}}^X(t,v_i^X)
 +I_i^{X,\mathrm{ext}}(t),
 \label{eq:fhn_voltage_dynamics}
\end{equation}
where \(F_{X,v}\) denotes the voltage component of the intrinsic FHN vector field.

\subsection{Hierarchy of intrinsic models}
The five models differ in state dimension and spike mechanism (Table~\ref{tab:model_summary}). LIF and EIF neurons evolve in a one-dimensional voltage space until they cross an imposed threshold, enter a \textit{refractory} state, and are subsequently reset. FHN and reduced HH neurons follow continuous nonlinear trajectories: an upward crossing of a prescribed detection surface records a spike but does not reset the state. In the FHN formulation used here, detected spikes are readouts only and do not generate recurrent coupling events; in the reduced HH networks, detected spikes do generate synaptic events. The model equations and model-specific information are collected in Appendix~\ref{app:model_details}.

For LIF neurons, the intrinsic drift is linear. EIF neurons add an exponential spike-initiation term while retaining the threshold--refractory--reset rule. An FHN neuron has a fast voltage-like coordinate and a slow recovery coordinate. The reduced HH-2D model combines voltage with one effective gating coordinate, whereas HH-3D retains potassium activation and sodium inactivation coordinates while treating sodium activation as instantaneous. This hierarchy tests the surrogate as phase-space dimension increases and as spike generation changes from imposed resets to continuous trajectories.

\begin{table*}[htbp]
\caption{Neuronal models considered in this study. A detection crossing is an upward crossing of the model-specific spike surface.}
\label{tab:model_summary}
\begin{ruledtabular}
\begin{tabular}{lcccc}
Model & State & Dimension & Spike event & Recurrent consequence \\
\hline
LIF & $V_i^X$ & 1 & Threshold, refractory state, and reset & Delivered synaptic event \\
EIF & $V_i^X$ & 1 & Threshold, refractory state, and reset & Delivered synaptic event \\
FHN & $(v_i^X,w_i^X)$ & 2 & Continuous excursion; detection crossing & Voltage differences \\
HH-2D & $(V_i^X,w_i^X)$ & 2 & Continuous action potential; detection crossing & Delivered synaptic event \\
HH-3D & $(V_i^X,n_i^X,h_i^X)$ & 3 & Continuous action potential; detection crossing & Delivered synaptic event \\
\end{tabular}
\end{ruledtabular}
\end{table*}

\section{Markovian surrogate and firing-rate estimators}
\label{sec:markov}

\subsection{Multidimensional phase-space discretization}

For population $X$, let $\mathcal P_X=\{C_\alpha^X:\alpha\in\Gamma_X\}$ be a finite partition of $\Omega_X$. For LIF and EIF models, we augment the index set by a refractory state, $\widehat\Gamma_X=\Gamma_X\cup\{\mathcal R_X\}$; otherwise $\widehat\Gamma_X=\Gamma_X$. We write the population state as a normalized reduced-model occupancy row vector $\bm\rho_t^X$ with components
\begin{equation}
 \rho_{t,\alpha}^X\ge0,
 \qquad
 \sum_{\alpha\in\widehat\Gamma_X}\rho_{t,\alpha}^X=1.
 \label{eq:population_distribution}
\end{equation}
Here $\rho_{t,\alpha}^X$ is the reduced mass assigned to cell $C_\alpha^X$. Thus, the one-dimensional voltage occupancy used for LIF and EIF becomes a normalized distribution over a $d_X$-dimensional grid for FHN and HH neurons.

For the pending-event LIF and EIF constructions, given the instantaneous mean synaptic drive $\bar{\bm H}_t^X=(\bar H_t^{XE},\bar H_t^{XI})$, the finite-state generator is
\begin{equation}
 Q_X(\bar{\bm H}_t^X)=D_X(I_X^0)+\lambda^{X,\mathrm{ext}}J_X^{\mathrm{ext}}+\sum_{Y\in\{E,I\}}\frac{\bar H_t^{XY}}{\tau^{XY}}J_X^{XY}+R_X.
 \label{eq:generator}
\end{equation}
Here $D_X(I_X^0)$ discretizes the full continuous drift, including the homogeneous constant current and any recurrent current represented continuously, $J_X^{\mathrm{ext}}$ and $J_X^{XY}$ describe the cell-to-cell redistribution caused by one external or recurrent input event, and $R_X$ releases an integrate-and-fire neuron from $\mathcal R_X$ to its reset cell. Operators that do not apply to a model are set to zero. For LIF and EIF, threshold-crossing transitions encoded by $D_X$ or a jump matrix are redirected into $\mathcal R_X$; therefore, firing transitions inherit the rate of the drift or input event that causes them and are not added as a separate, drive-independent block. The FHN and HH generators have different model-specific continuous-transport terms: FHN depends on the population mean voltages, whereas HH depends on the mean synaptic activations. Their explicit forms are given in Appendix~\ref{app:discretization}. 

The variables $H_i^{XY}$ represent pending recurrent input events. A delivered type-$Y$ spike increments the corresponding pool of a type-$X$ neuron, and each pending event is consumed after an exponentially distributed waiting time with mean $\tau^{XY}$, at which point the jump represented by $J_X^{XY}$ is applied. This delayed-event construction has the same mean exponential impulse response as Eq.~\eqref{eq:synaptic_kernel}. Instantaneous external inputs are represented directly by the Poisson event rate $\lambda^{X,\mathrm{ext}}$ and $J_X^{\mathrm{ext}}$.

Equation~\eqref{eq:generator} therefore uses the instantaneous-external-input convention. For the FHN formulation, the electrical-coupling coefficients $G_{\mathrm{gap}}^{XY}$ are time independent and enter the continuous transport operator $D_X$. Detected spikes are readouts, so the recurrent pending-event terms and their mean-drive equations are absent. The generator therefore specializes by model, retaining only the applicable transport and event operators. 

The population distribution follows the Kolmogorov forward equation
\begin{equation}
 \frac{d\bm\rho_t^X}{dt}=\bm\rho_t^XQ_X(\bar{\bm H}_t^X).
 \label{eq:forward}
\end{equation}
To close the population equations, the pending recurrent inputs are replaced by their expectations. This step assumes that the pending drive of a neuron decorrelates rapidly from the neuron's past activity and that neurons of the same population occupying the same state cell are exchangeable \cite{chang2025minimizing}. In the earlier LIF formulation, annealed connectivity motivates this factorization by suppressing persistent correlations between a neuron’s state and its realized input pool. Under this closure,
\begin{equation}
 \frac{d\bar H_t^{XY}}{dt}=-\frac{\bar H_t^{XY}}{\tau^{XY}}
 +c^{XY}f_t^Y.
 \label{eq:mean_drive}
\end{equation}
Here $c^{XY}$ is the expected number of delivered type-$Y$ contacts onto one type-$X$ neuron. Excluding autapses, $c^{XY}=(N_Y-\mathds{1}_{X=Y})P^{XY}$.

Equations~\eqref{eq:forward} and \eqref{eq:mean_drive} constitute a first-moment closure. They retain temporal modulation of the mean synaptic drive but do not explicitly evolve variances, pair correlations, or higher-order moments.

\subsection{A model-independent firing-rate flux}

Let $\mathcal E_X^{\mathrm{fire}}$ be the set of directed cell transitions that count as spikes: transitions into $\mathcal R_X$ for LIF and EIF, and upward crossings of the prescribed detection surface for FHN and HH. In this study, the detection surface for FHN and HH has the form $S^X=\{(V^X,u^X_2,\cdots,u^X_{d_X})|V^X=v^{X,\mathrm{th}}\}$; a directed cell transition counts as a spike when it crosses this surface upward, from a state with $V^X<v^{X,\mathrm{th}}$ to $V^X\ge v^{X,\mathrm{th}}$. 

For $\alpha\ne\beta$, define the spike-counting matrix
\begin{equation}
 [K_X(\bar{\bm H})]_{\alpha\beta}
 =[Q_X(\bar{\bm H})]_{\alpha\beta}
 \mathds{1}_{\{(\alpha,\beta)\in\mathcal E_X^{\mathrm{fire}}\}},
 \qquad [K_X]_{\alpha\alpha}=0.
 \label{eq:spike_matrix}
\end{equation}
The instantaneous population firing rate is the probability flux through these transitions,
\begin{equation}
 f_t^X=\bm\rho_t^XK_X(\bar{\bm H}_t^X)\mathbf{1},
 \label{eq:firing_flux}
\end{equation}
where $\mathbf{1}$ denotes the all-ones column vector of the appropriate dimension.
For an integrate-and-fire surrogate with one refractory state and exponential release rate $1/\tau_{\mathrm{ref}}^X$, stationarity gives the equivalent identity
\begin{equation}
 f_\infty^X=\frac{\rho_{\infty,\mathcal R_X}^X}{\tau_{\mathrm{ref}}^X}.
 \label{eq:refractory_flux}
\end{equation}
We comment that Eq.~\ref{eq:refractory_flux} is equivalent to an exponentially distributed refractory time with mean $\tau_{\mathrm{ref}}^X$. This does not exactly follow the classical setup in LIF and EIF models, but is rather a sacrifice to ensure the Markovian property. The author used similar setup in \cite{cai2021model,chang2025minimizing}.
On the other hand, Eq.~\eqref{eq:firing_flux} also applies when no refractory state exists.

\subsection{Closures for two estimators}

The \emph{Type I estimator} seeks a time-independent self-consistent solution,
\begin{align}
 \bm{0}&=\bm\rho_\infty^XQ_X(\bar{\bm H}^X),
 &1&=\bm\rho_\infty^X\mathbf{1},
 \label{eq:type1_distribution}\\
 0&=-\frac{\bar H^{XY}}{\tau^{XY}}+c^{XY}f_\infty^Y,
 &f_\infty^X&=\bm\rho_\infty^XK_X(\bar{\bm H}^X)\mathbf{1}.
 \label{eq:type1_closure}
\end{align}
Although the null-vector problem is linear for fixed $\bar{\bm H}$, the coupled self-consistency problem is generally nonlinear. The Type I estimator enforces stationarity of both $\bm\rho^X$ and $\bar{\bm H}^X$.

The \emph{Type II estimator} integrates Eqs.~\eqref{eq:forward}, \eqref{eq:mean_drive}, and \eqref{eq:firing_flux}. Its predicted firing rate is the long-time average
\begin{equation}
 \widehat f^X=\frac{1}{T_1-T_0}\int_{T_0}^{T_1}f_t^X\,dt,
 \label{eq:type2_average}
\end{equation}
where $[0,T_0]$ is discarded as a transient. Type II can therefore represent synaptic-timescale effects that disappear from the stationary mean-drive balance. Refractory-period changes affect both estimators through refractory occupancy, and in Type II they may additionally alter the coupled temporal dynamics. It does not, by construction, resolve neuron-to-neuron fluctuations or higher-order input moments. The reported numerical integration uses the forward-Euler method; Appendix~\ref{app:numerics} gives the numerical settings.

Neither closure simulates the spike sequence of every neuron in the full network. Instead, Type I solves a finite-state fixed point and Type II integrates a deterministic reduced system. Because the number of phase-space cells grows rapidly with $d_X$, the three-dimensional HH example serves as a feasibility test.

\section{Results}
\label{sec:results}

We begin with a controlled LIF benchmark and then increase the nonlinearity and dimension of the intrinsic neuronal dynamics.

\subsection{Detailed LIF benchmark}
\label{sec:lif_results}

We first use the LIF model to test the two closures in detail. Direct LIF simulations provide the numerical reference values. We use the spike synchrony index (SSI) as a diagnostic of coordinated population activity \cite{chariker2018rhythm}. Let $t_k$ be the time of the $k$th population spike, let $s(k)$ be its source neuron, and let ${t_{j,\mu}}$ be the spike times of neuron $j$. We define
\[
\text{SSI}=\frac{1}{N_{\text{spikes}}}\sum_{k=1}^{N_{\text{spikes}}}\frac{1}{N}\sum_{j\neq s(k)}\sum_\mu\mathds{1}_{|t_{j,\mu}-t_k|<\frac{w}{2}},
\]
with $w=5~\mathrm{ms}$.

\begin{figure}[tbp]
 \centering
 \includegraphics[width=\linewidth]{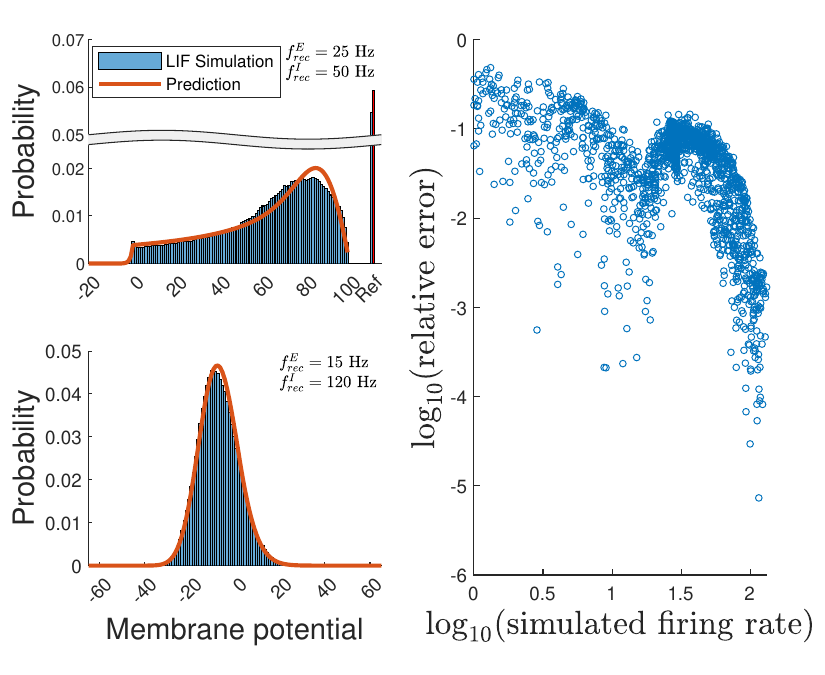}
 \caption{The Type I estimator for a single LIF neuron receiving temporally homogeneous Poisson input. Left: two examples compare the long-time simulated voltage histogram with the Markovian stationary prediction; the narrow bar marked ``Ref'' is the refractory-state probability. Right: relative firing-rate errors for 3,000 parameter configurations plotted against the simulated firing rates on logarithmic axes. Recurrent input rates, refractory periods, and external input rates are varied.}
 \label{fig:lif_single}
\end{figure}

For a single LIF neuron receiving independent, constant-rate excitatory and inhibitory Poisson inputs, the recurrent rates are prescribed. Type I then reduces to a null-vector problem for $Q_X$. Figure~\ref{fig:lif_single} compares its predicted stationary voltage distribution and firing rate with the corresponding quantities from long-time direct simulations of duration $T=100$~s and step size $\delta t=0.1$~ms. Across 3,000 configurations, the recurrent rates range from 0 to 75~Hz, refractory periods from 0.8 to 2.3~ms, and external input rates from 100 to 10,000~Hz. In both presented examples, the predicted voltage distribution closely matches the corresponding simulated histogram, and the median relative firing-rate error over the full sweep is 7.04\%. The relative error generally decreases as the simulated rate increases; at very low rates, normalization by a small reference value makes the relative error particularly sensitive to small absolute discrepancies.

\begin{figure}[tbp]
 \centering
 \includegraphics[width=0.99\linewidth]{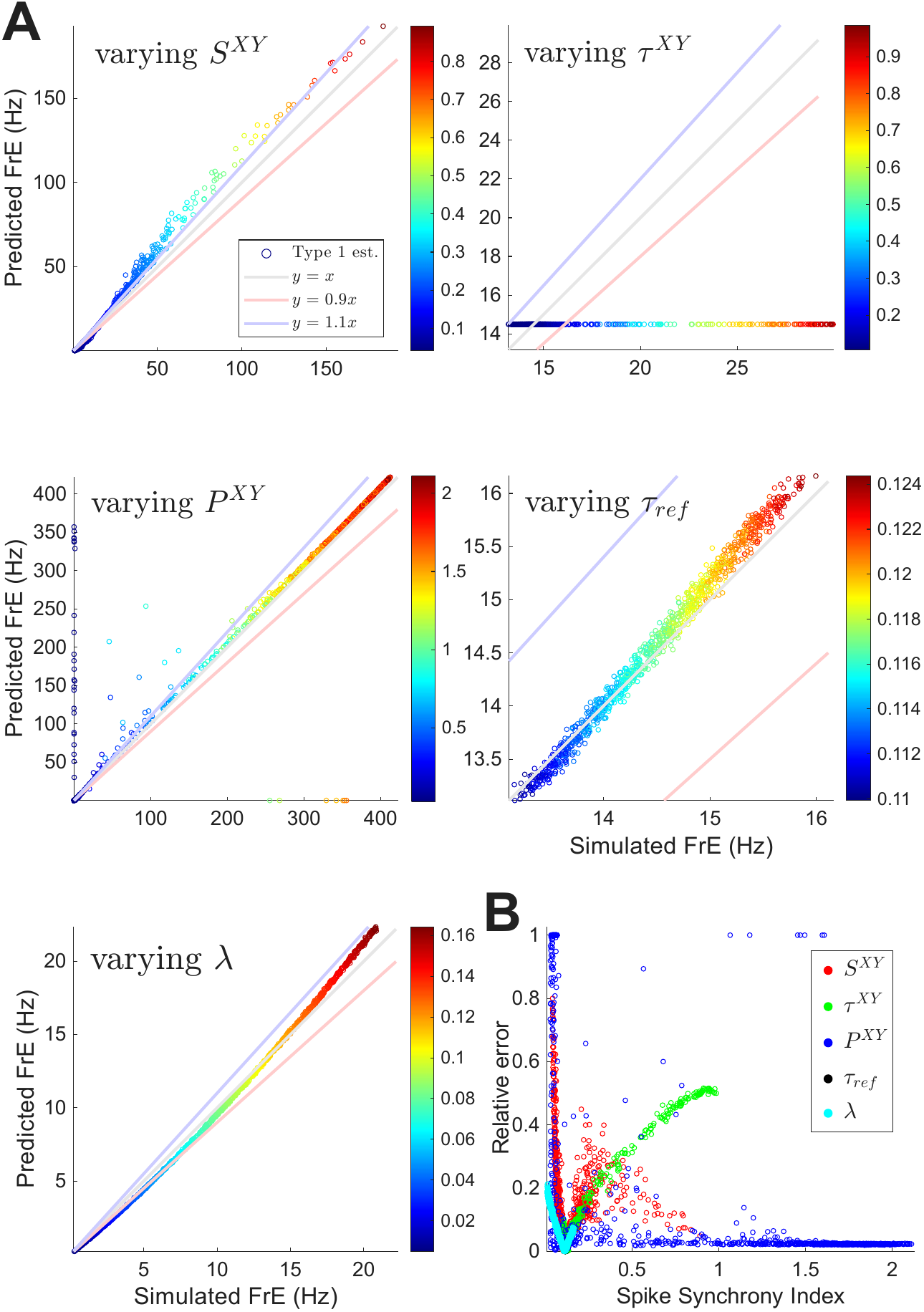}
 \caption{The Type I estimator for recurrent LIF networks with 300 excitatory and 100 inhibitory neurons. (A) Predicted versus simulated excitatory firing rates (FrE) for 5,000 parameter sets. Separate panels vary recurrent strengths $S^{XY}$, synaptic timescales $\tau^{XY}$, coupling probabilities $P^{XY}$, refractory periods $\tau_{\mathrm{ref}}^X$, and external rates $\lambda^{X,\mathrm{ext}}$; color encodes SSI. The diagonal and the two flanking lines denote equality and $\pm10\%$, respectively. (B) Relative error versus SSI, with symbols grouped by the varied parameter family.}
 \label{fig:lif_type1}
\end{figure}

We next solve the Type I self-consistency problem for recurrent networks with 300 excitatory and 100 inhibitory neurons. The 5,000 tested parameter sets vary recurrent strengths, coupling probabilities, external rates, synaptic timescales, and refractory periods over the ranges in Table~\ref{tab:lif_parameters}. For the $S^{XY}$, $P^{XY}$, $\tau_{\text{ref}}^X$, and $\lambda^{X,\mathrm{ext}}$ sweeps, most predictions follow the simulated rates, apart from very low-rate cases and the abrupt activity changes discussed in Appendix~\ref{app:branches}. In contrast, the Type I estimator produces nearly unchanged rates when $\tau^{XY}$ is varied, although the corresponding simulations span a much wider rate range. The stationary closure therefore cannot reproduce the observed timescale-dependent rate variation.

The relationship between SSI and Type I error is parameter dependent (Fig.~\ref{fig:lif_type1}B). The synaptic-timescale cases show substantial errors over part of the SSI range. Most low-SSI outliers coincide with nearly zero simulated firing, for which the relative-error metric is ill-conditioned (see the low-firing-rate cases in the $S^{XY}$ and $P^{XY}$ panels of Fig.~\ref{fig:lif_type1}A). 

\begin{figure}[tbp]
 \centering
 \includegraphics[width=0.99\linewidth]{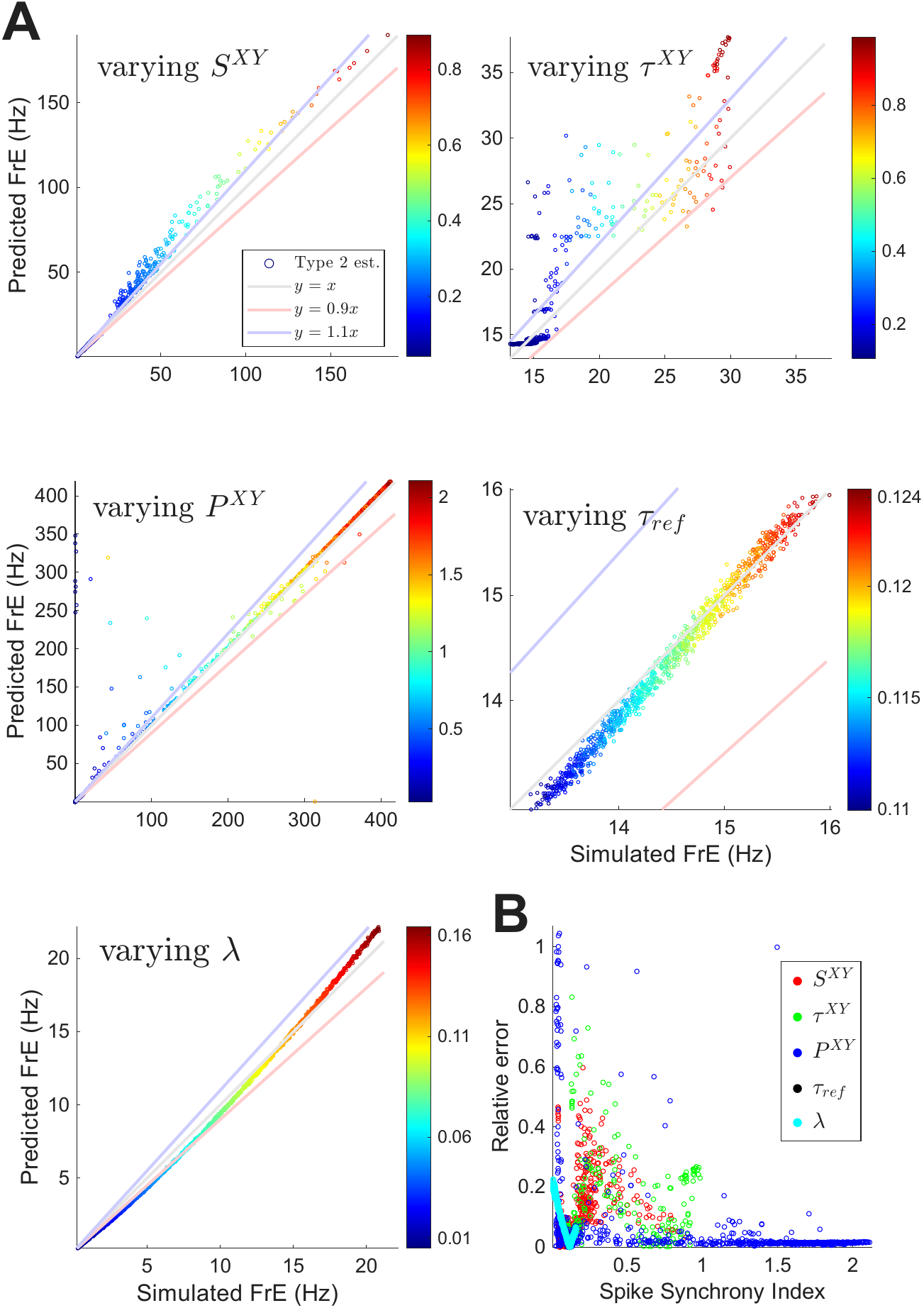}
 \caption{The Type II estimator for the same recurrent LIF networks and 5,000 parameter sets as in Fig.~\ref{fig:lif_type1}. (A) Predicted versus simulated excitatory firing rates for the five parameter families; color encodes SSI. (B) Relative error versus SSI. Retaining the temporal evolution of the mean synaptic drive substantially improves accuracy in the synaptic-timescale sweep, although outliers remain.}
 \label{fig:lif_type2}
\end{figure}

The Type II estimator retains the temporal mean-drive feedback and responds to both synaptic timescales and refractory periods (Fig.~\ref{fig:lif_type2}). For the $\tau^{XY}$ sweep, more than 84\% of cases have relative errors below 10\%, and more than 92\% have errors below 30\%. Performance in the other sweeps remains parameter dependent, with visible low-rate and coupling-probability outliers. These results are consistent with temporal mean-drive dynamics being an important component missing from the stationary closure in Type I, though higher-order statistics of drive are still ignored.

\begin{figure}[htbp]
 \centering
 \includegraphics[width=0.99\linewidth]{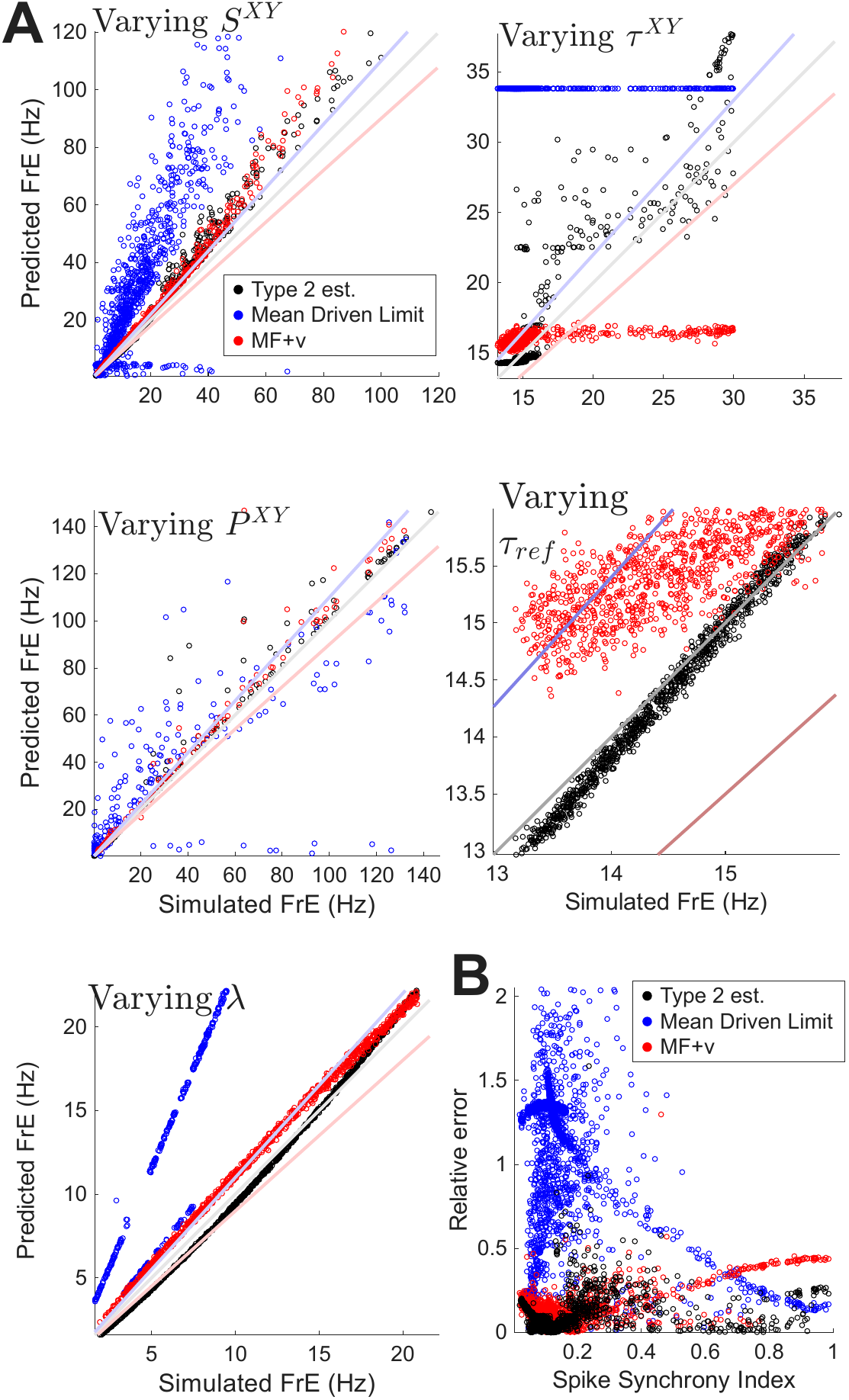}
 \caption{Prediction error comparison for the Type II estimator, the mean-driven limit, and MF+$v$ in LIF networks with 300 excitatory and 100 inhibitory neurons. (A) Predicted versus simulated excitatory firing rates while $S^{XY}$, $\tau^{XY}$, $P^{XY}$, $\tau_{\mathrm{ref}}^X$, and $\lambda^{X,\mathrm{ext}}$ are varied; the diagonal denotes equality. The mean-driven limit is not applied to the refractory-period sweep. (B) Relative error versus SSI.}
 \label{fig:lif_compare}
\end{figure}

Figure~\ref{fig:lif_compare} compares the Type II estimator with the mean-driven limit \cite{cai2006kinetic,treves1993meanfield} and the data-informed mean-field-plus-voltage (MF+$v$) method \cite{xiao2021data}. The version of the mean-driven approximation used here omits synaptic-timescale dynamics and does not include a nonzero refractory period; it is therefore not shown for the refractory-period sweep. Its predictions are consequently insensitive to $\tau^{XY}$ and show large deviations in parts of the recurrent-strength sweep. MF+$v$ incorporates simulated single-neuron responses and is closer than the mean-driven limit in many displayed cases, but it does not represent collective temporal modulation. Within the sampled regimes, the Type II estimator often lies closer to the equality line than the alternatives, especially where synaptic-drive dynamics matter. 


\subsection{EIF robustness test}
\label{sec:eif_results}

The EIF model tests whether the one-dimensional construction remains useful when the subthreshold voltage drift contains a nonlinear exponential term. Because the EIF results qualitatively resemble the LIF results, we summarize them here and place the corresponding figures in Appendix~\ref{app:eif_results} (Figs.~\ref{fig:eif_single}--\ref{fig:eif_compare}). 

In the two single-neuron examples in Fig.~\ref{fig:eif_single}, the stationary Markovian voltage profiles follow the corresponding long-time simulation histograms. In one example, probability is concentrated at subthreshold voltages, whereas the other has appreciable refractory occupancy. Across the displayed single-neuron sweep, larger relative errors occur primarily at low firing rates, whereas many higher-rate cases have substantially smaller relative errors.

Within the reported recurrent EIF sweeps, the comparison in Figs.~\ref{fig:eif_type1} and \ref{fig:eif_type2} has the same qualitative structure as the LIF benchmark. The Type I estimator is insensitive to the synaptic-timescale sweep, while the Type II estimator produces a wider range of predictions and captures part of the simulated variation. Both estimators show large deviations in subsets of the coupling-probability and timescale sweeps. Figure~\ref{fig:eif_compare} further shows that the Type II estimator tracks the reference rates more closely than the mean-driven limit and MF+$v$ in many of the displayed cases, although the errors of all three methods remain parameter dependent. 

Before proceeding to network examples composed of multidimensional single neurons, we systematically compare the computation cost of our methods to classical methods and direct network simulations in Table~\ref{table:runtimes_1d}.  The requested time window to control rate estimate error could be long as 100 seconds for network dynamics with large Fano factors and high synchrony between neurons (see Appendix). Nevertheless, we use ten second as our baseline simulation time window for networks, since it approximately controls the rate estimate error to 20\% for a single Poissonian neuron with 10Hz firing rate. In general, the Type I estimator can speed up the firing rate estimates for multiple magnitudes, while the Type II estimator sacrifices speed for more accuracy.

\begin{table*}[htbp]
\begin{tabular}{|c|c|c|c|c|c|c|}
\hline
Model                & $N$    & Direct simulation (10~s) & Type I   & Type II  & MF+$v$      & Mean-driven limit \\ \hline
LIF & $4$    & $5.3721$          & $0.0184$ & $4.8634$ & $1.3220$  & $0.0004$          \\ \cline{2-7} 
                     & $40$   & $5.1988$          & $0.0209$ & $4.1990$ & $2.2439$  & $0.0005$          \\ \cline{2-7} 
                     & $400$  & $8.6068$          & $0.0229$ & $4.7958$ & $7.5157$  & $0.0006$          \\ \cline{2-7} 
                     & $4000$ & $55.9751$         & $0.0221$ & $7.2409$ & $61.0321$ & $0.0007$          \\ \hline
EIF & $4$    & $2.7125$          & $0.0607$ & $6.1643$ & $1.3936$  & $0.0014$          \\ \cline{2-7} 
                     & $40$   & $3.5690$          & $0.0711$ & $6.4884$ & $2.8546$  & $0.0014$          \\ \cline{2-7} 
                     & $400$  & $8.3846$          & $0.2002$ & $7.6239$ & $8.1269$  & $0.0013$          \\ \cline{2-7} 
                     & $4000$ & $62.9410$         & $0.1994$ & $7.6137$ & $62.0735$ & $0.0011$          \\ \hline
\end{tabular}
\caption{Mean wall-clock runtime (seconds) of the one-dimensional algorithms, including direct network simulations, both types of estimators, MF+$v$, and the mean-driven limit. Values are averages over 1,000 parameter sets for $N=4$, $40$, and $400$, and over 50 parameter sets for $N=4{,}000$. Each network simulation represents 10~s of biological time with $\delta t = 0.1~\mathrm{ms}$.}
\label{table:runtimes_1d}
\end{table*}

\subsection{Two-dimensional phase-space distributions}
\label{sec:2d_distributions}

\begin{figure*}[htbp]
 \centering
 \includegraphics[width=1\textwidth]{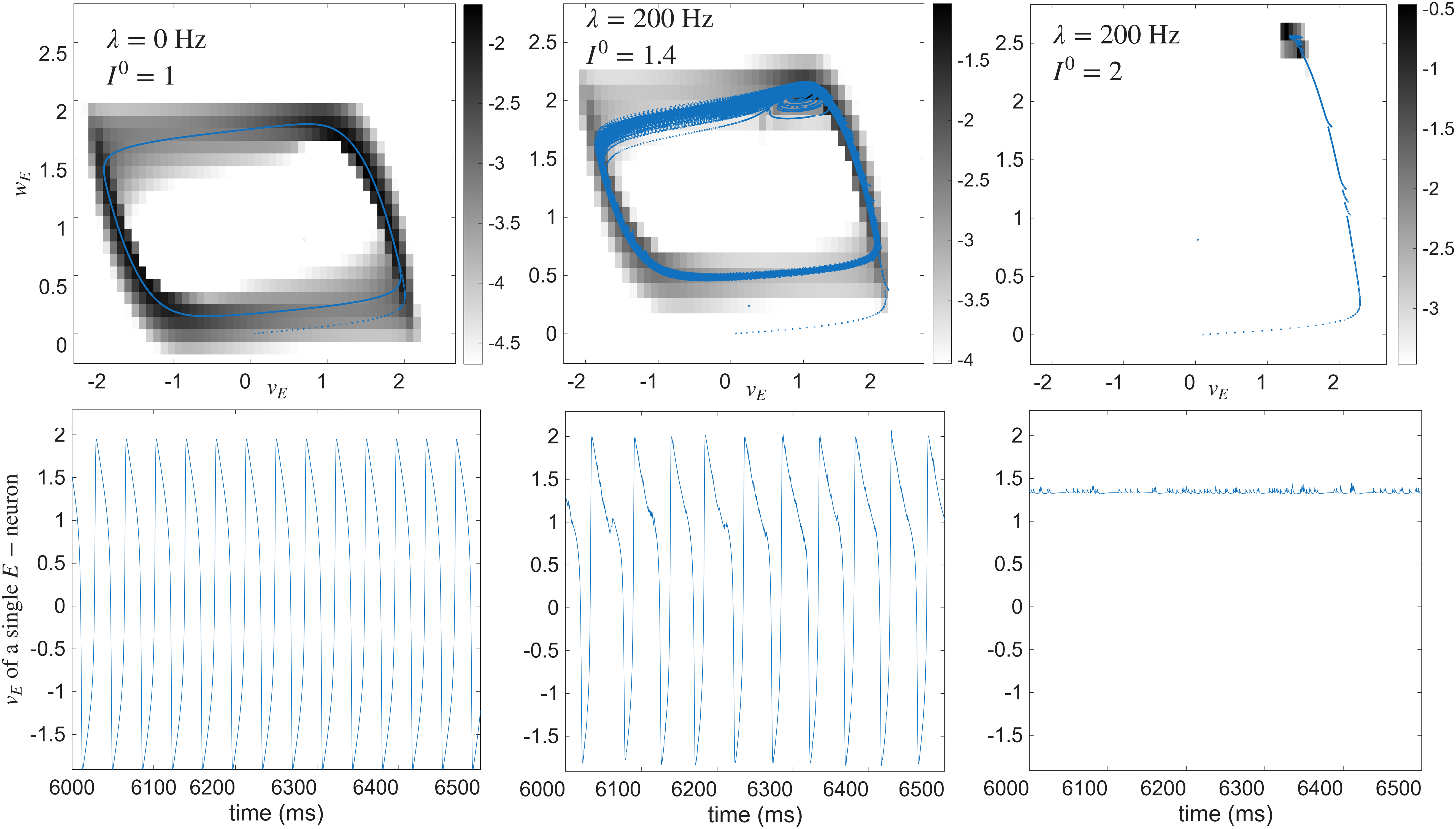}
 \caption{Predicted phase-space probabilities and representative simulated trajectories for the excitatory population of the FitzHugh--Nagumo (FHN) network under three parameter settings. In the upper row, grayscale represents $\log_{10}\rho_{\infty,\alpha}^E$, where $\rho_{\infty,\alpha}^E$ is the stationary probability mass assigned by the Type I estimator to cell $C_\alpha^E$. Darker shading denotes greater predicted probability. The blue curve in each upper panel shows the trajectory of one excitatory neuron from a direct simulation of a network with $N=400$ neurons (simulated for $7.5\,\mathrm{s}$ using a time step of $\delta t=0.05\,\mathrm{ms}$). The lower row shows the corresponding simulated voltage trace $v_E(t)$ over the displayed time interval.}
 \label{fig:fhn_distributions}
\end{figure*}

The FHN and HH-2D tests assess a different target: the long-time probability distribution over a two-dimensional phase space. Figure~\ref{fig:fhn_distributions} shows three FHN regimes. In the two oscillatory examples, occupancy is distributed nonuniformly around an extended loop; in the third, it is concentrated in a localized region and the voltage-like trace remains near a steady level. The overlaid paths pass through the occupied regions and include low-occupancy approach segments. These panels illustrate the target of the multidimensional surrogate: a long-time distribution supported along nonlinear trajectories rather than a one-dimensional voltage axis.

\begin{figure*}[htbp]
 \centering
 \includegraphics[width=1\textwidth]{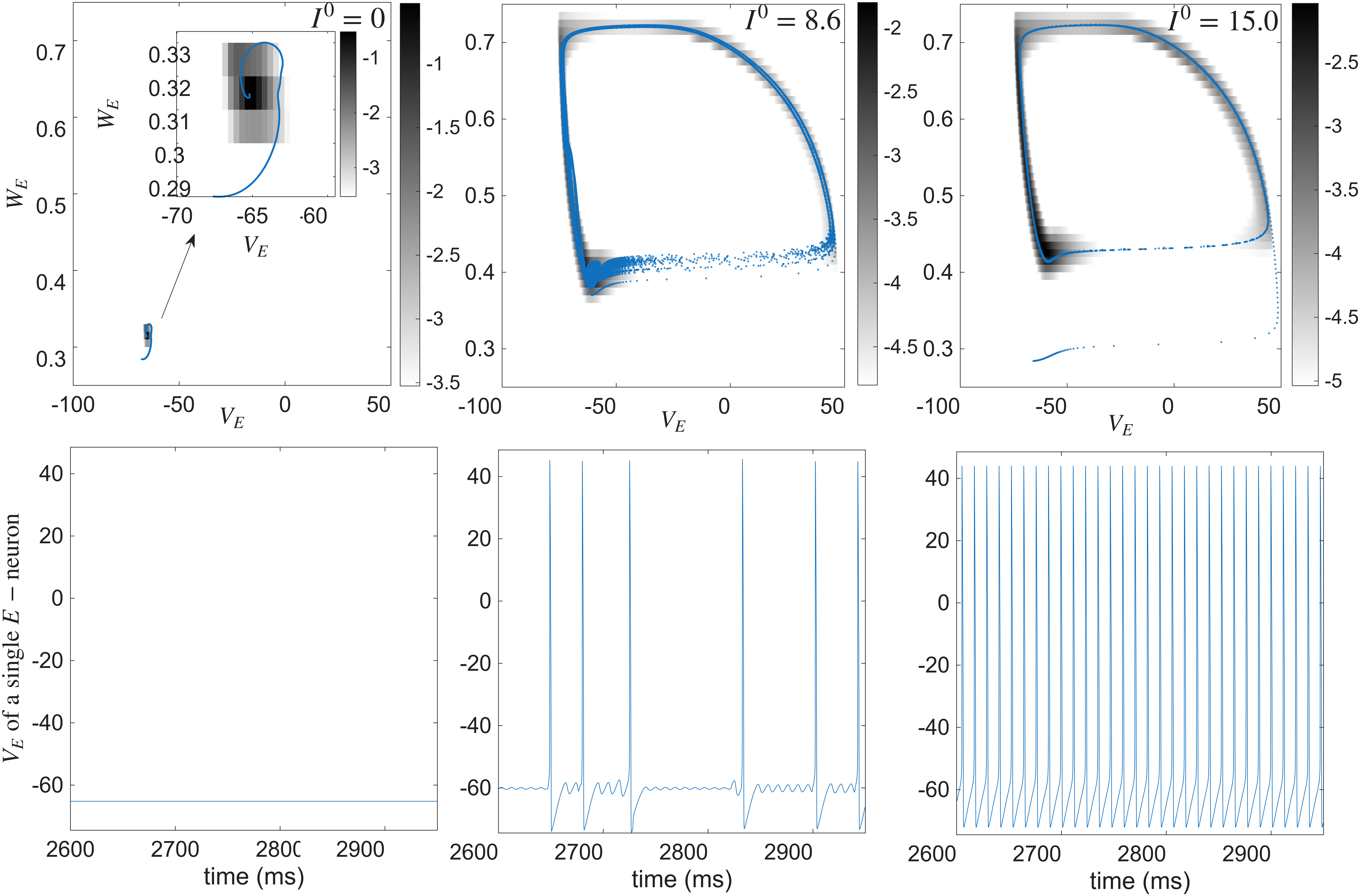}
 \caption{Predicted phase-space probabilities and representative simulated trajectories for the excitatory population of the 2D Hodgkin-Huxley (HH-2D) network under three parameter settings. Greyscales and blue trajectories in the upper and lower rows are similar to Fig.~\ref{fig:fhn_distributions}.}
 \label{fig:hh2d_distributions}
\end{figure*}

Figure~\ref{fig:hh2d_distributions} gives the analogous display for HH-2D. The columns progress from occupancy localized near a resting state, through sparse spike excursions, to repeated traversal of a large action-potential loop. In each column, the phase-space support changes consistently with the corresponding voltage trace. Together, the panels show how the target phase-space support changes across quiescent, sparse-spiking, and repetitive-spiking regimes.

\subsection{Two-dimensional firing-rate comparisons}
\label{sec:2d_rates}

\begin{figure}[htbp]
 \centering
 \includegraphics[width=0.99\linewidth]{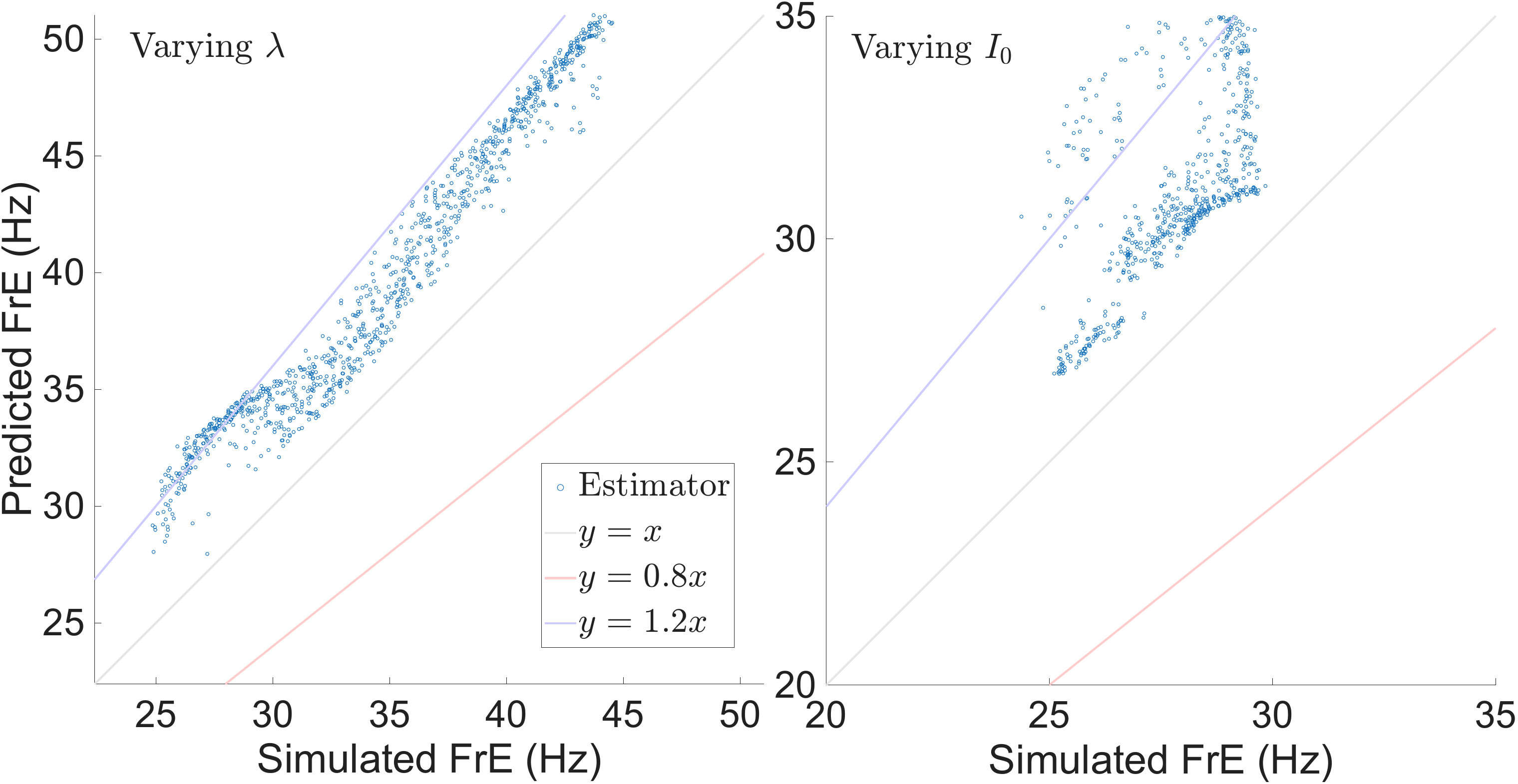}
 \caption{The Type I estimator versus simulated FHN firing rates. Each parameter sweep contains 1,000 FHN networks with 300 excitatory and 100 inhibitory neurons. Left: external Poisson rate $\lambda$. Right: constant external current $I_0$. The diagonal denotes equality, and the flanking lines denote 0.8 and 1.2 times the simulated rate.}
 \label{fig:fhn_rates}
\end{figure}

The FHN firing-rate sweeps in Fig.~\ref{fig:fhn_rates} show that the surrogate captures the variation in firing rate as external Poisson input rate $\lambda$ and external input current $I_0$ vary, but most displayed predictions lie above the equality line. Here we are testing gap-junction type coupling between neurons, i.e., the recurrent drive from neuron $i$ to $j$ relies on $(v_j-v_i)$ rather than exact spiking timings.

We find that the bias decreased under grid refinement $\delta v,\delta w\to 0$, consistent with numerical diffusion. Under a constant current input $I_0$, the discretization introduces an extra artificial variance in drive $\sim |f_v|\delta v$, pushing extra probability flow through the threshold. For other systems (LIF and EIF) in this study, the external drive is purely Poisson, hence the variance-induced flow around the threshold does not appear.

Because the FHN networks considered here are all-to-all connected with electrical couplings, the recurrent input can be expressed in terms of deviations from the population-mean voltages:
\begin{equation}
J_{i,\mathrm{gap}}^X
=
G_{\mathrm{gap}}^{XE}N_E\left(\bar v_E-v_i^X\right)
+
G_{\mathrm{gap}}^{XI}N_I\left(\bar v_I-v_i^X\right).
\label{eq:fhn_gap_mean}
\end{equation}
Thus, the recurrent input is provided by voltage differences between individual neurons and the two population means. In the long-time regimes examined here, neurons subject to statistically identical external forcing remained close to their respective population means, with $\bar v_E\approx\bar v_I$, so both differences in Eq.~\eqref{eq:fhn_gap_mean} were typically small. Independent Poisson inputs prevent exact synchronization, and the coupling term therefore does not vanish identically. Nevertheless, numerical checks (not shown) over the examined range $(0,0.002)$ of $G_{\mathrm{gap}}^{XY}$ produced only minor changes in the mean firing rates. We therefore restricted the reported parameter sweeps to the external Poisson rate $\lambda$ and constant current $I_0$, which directly modify the stochastic and deterministic drives, respectively. These couplings may still affect synchronization, transient relaxation, and finite-size fluctuations, which are not studied in this paper.

\begin{figure}[htbp]
 \centering
 \includegraphics[width=0.99\linewidth]{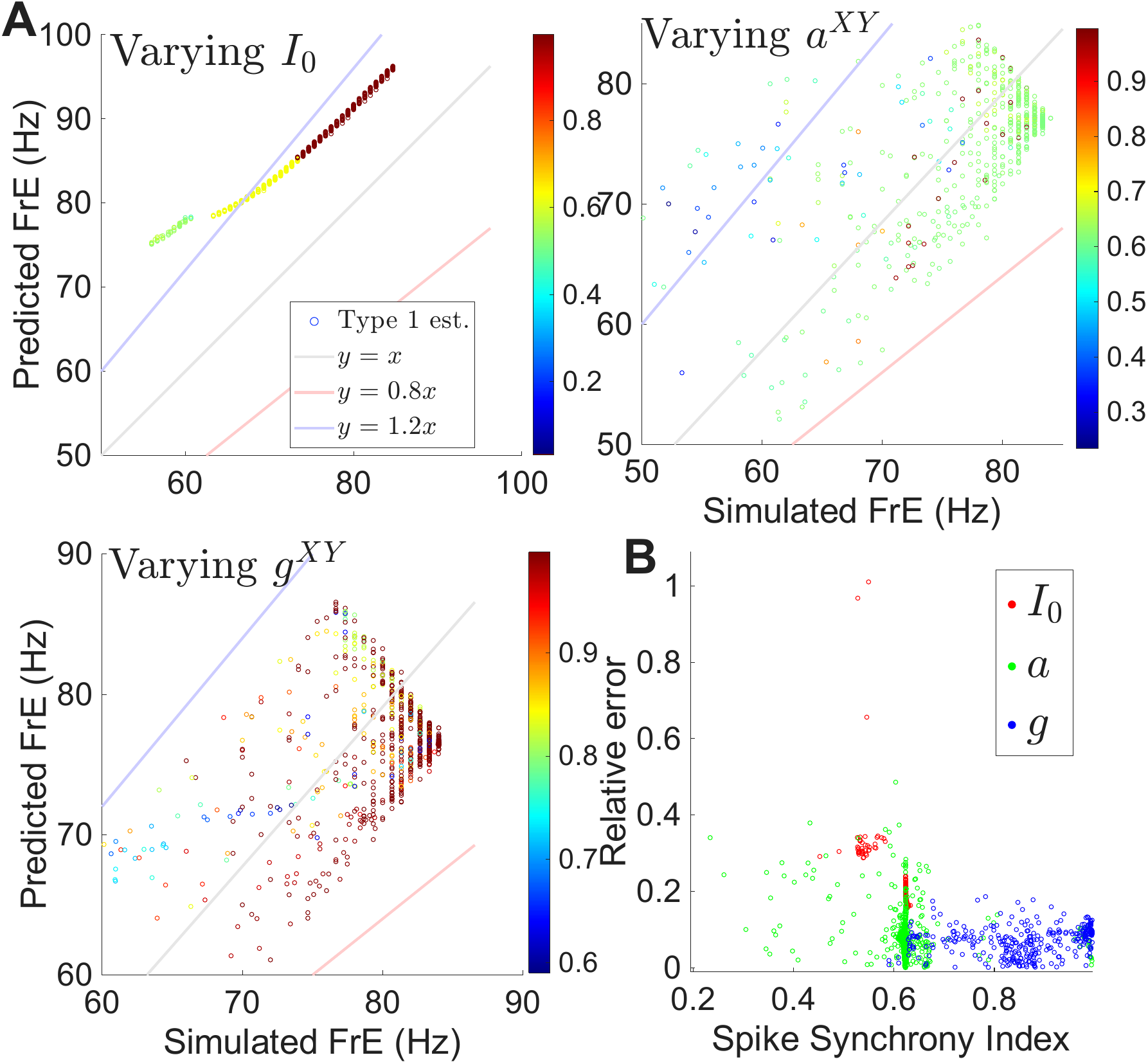}
 \caption{The Type I estimator for HH-2D networks with 300 excitatory and 100 inhibitory neurons. (A) Predicted versus simulated rates while constant input current $I_0$, $a^{XY}$, and $g^{XY}$ are varied; the diagonal denotes equality and the flanking lines denote $0.8$ and $1.2$ times the simulated rate. (B) Relative error versus SSI, grouped by parameter family. Accuracy is parameter dependent, with a systematic upward bias in the $I_0$ sweep and broader scatter in the $a^{XY}$ and $g^{XY}$ sweeps.}
 \label{fig:hh2d_rates}
\end{figure}

For HH-2D, the predicted rates vary smoothly with $I_0$, but the corresponding points lie systematically above the equality line (Fig.~\ref{fig:hh2d_rates}). The $a^{XY}$ and $g^{XY}$ sweeps show broader scatter, with both under- and overestimation; many $g^{XY}$ cases lie near equality, but visible outliers remain. 
Together, the FHN and HH-2D panels extend the firing-rate comparison beyond threshold--reset models and reveal bias patterns that differ from those in the LIF benchmark.

The computation cost of our methods is compared to direct network simulation in Table~\ref{table:runtimes_2d}. Due to the increasing demand in discretizations of 2D fields, the Type I estimator becomes more costly than the 1D case, but it is still not more expensive than simulating 400-neuron networks for 10 seconds. We stress again that a reasonable simulation time to control rate estimate error could be much longer than this due to irregular firing dynamics.

\subsection{Three-dimensional HH feasibility test}
\label{sec:hh3d_results}

\begin{figure}[htbp]
 \centering
 \includegraphics[width=\linewidth]{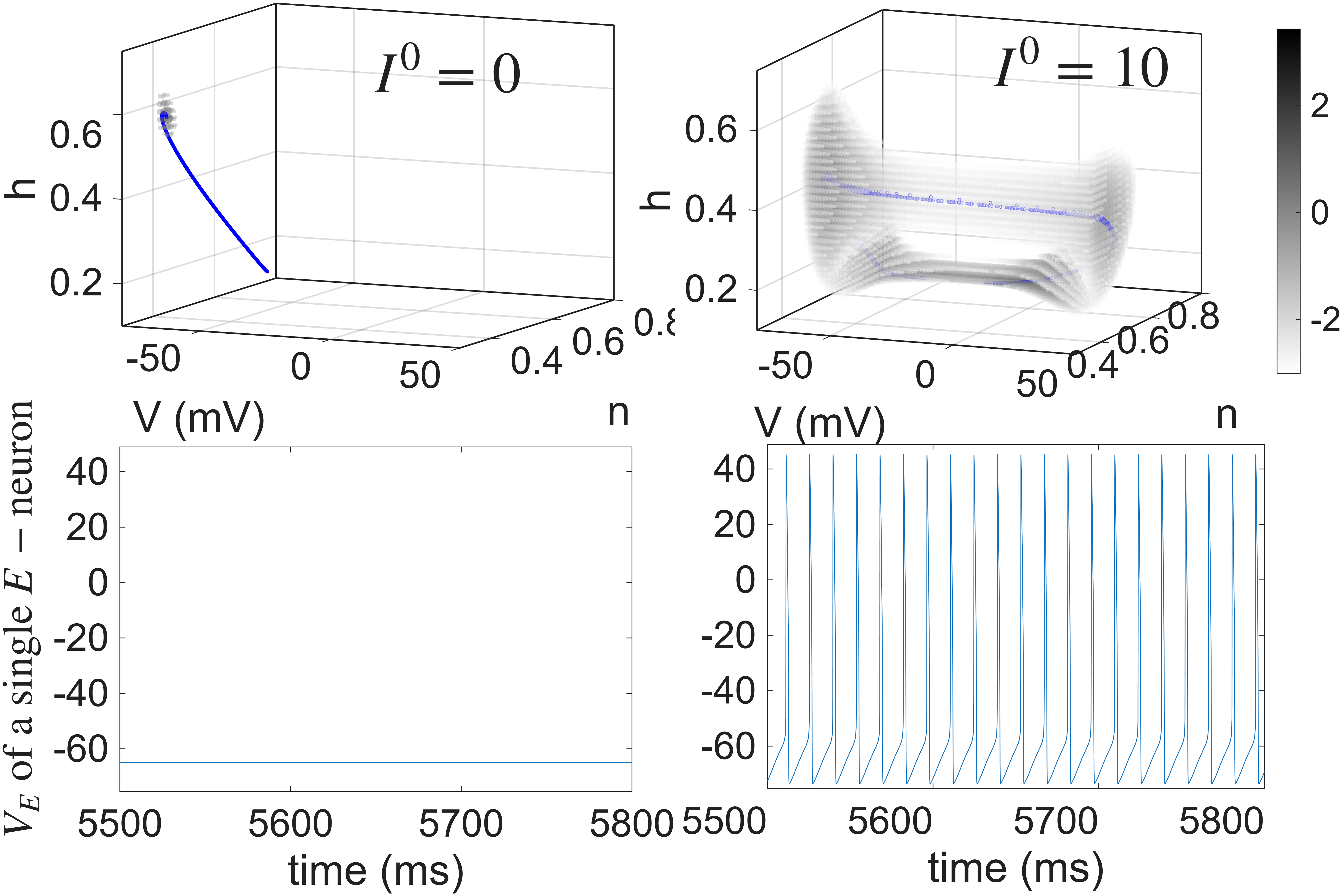}
 \caption{HH-3D phase-space occupancy for two regimes. Top: Type I occupancy and an overlaid path in $(V,n,h)$ space. Bottom: the corresponding voltage trace. In the left column, occupancy is localized near a quiescent region and the overlaid path includes an approach segment; in the right column, occupancy follows an extended loop accompanied by repeated spiking.}
 \label{fig:hh3d_distributions}
\end{figure}

The HH-3D examples in Fig.~\ref{fig:hh3d_distributions} establish that the finite-state construction can represent both localized and loop-like support in a three-dimensional intrinsic state space. They are intended as a feasibility test only: we did not conduct a systematic firing-rate or grid-convergence study in three dimensions.

\begin{table*}[htbp]
\begin{tabular}{|c|c|c|}
\hline
Model & Direct simulation (10~s) & Type I   \\ \hline
FHN   & $5.8498$          & $5.6333$  \\ \hline
HH-2D & $39.5511$         & $23.9955$ \\ \hline
\end{tabular}
\caption{Mean wall-clock runtime (seconds) for the FHN and HH-2D calculations. Values are averages over 1,000 parameter sets for networks of $N=400$ neurons. Each direct simulation represents 10~s of biological time and $\delta t = 0.1$~ms.}
\label{table:runtimes_2d}
\end{table*}

\section{Discussion and conclusions}
\label{sec:discussion}

We developed finite-state estimators of long-time state distributions and mean firing rates that avoid spike-by-spike simulation of the full network. In both the LIF benchmarks and the EIF tests, the stationary closure is insensitive to synaptic timescales, whereas evolving the mean pending input restores this dependence but does not eliminate all regime-dependent errors. The FHN and HH-2D examples show that the same occupancy-and-flux idea can be applied to continuous multidimensional spike trajectories, although their rate estimates retain systematic or parameter-dependent biases. The HH-3D example establishes representational feasibility in three dimensions. 

Compared with the mature theory for one-dimensional integrate-and-fire populations \cite{NykampTranchina2000,brunel2000dynamics,MontbrioEtAl2015}, quantitative rate reductions that retain several intrinsic neuronal state variables remain less developed and more model-specific. Existing approaches include mean-field limits for fully connected FHN and HH networks, population-density and low-dimensional reductions for adaptive integrate-and-fire neurons, transfer-function-based descriptions of conductance-based AdEx, HH, and Morris--Lecar networks, and numerical population-density methods for general multidimensional states \cite{baladron2012meanfield,nicola2013twodim,augustin2017lowdim,zerlaut2018mesoscopic,carlu2020complex,osborne2022ndimensional}. Our construction complements these approaches by using a common finite-state occupancy and firing-flux representation across models with one to three intrinsic dimensions. Relative to general multidimensional population-density solvers, our emphasis is therefore on a common finite-state representation together with model-specific recurrent feedback and spike detection.

Compared to a scalar rate closure, retaining the distribution of neuronal states provides access to the probability mass near the firing surface. Equation~\eqref{eq:firing_flux} writes firing as probability flux, $f_t^X=\bm\rho_t^XK_X(\bar{\bm H}_t^X)\mathbf{1}$; two populations with the same mean state and mean drive can therefore have different rates if their near-threshold mass differs. This phase-space advantage is shared with population-density and kinetic descriptions and is not unique to the present method. Here it is combined with jump matrices that move probability by finite synaptic events. Such events can produce rate responses missed by a diffusion approximation \cite{OmurtagEtAl2000,richardson2010shotnoise}. We note that in previous work on quadratic integrate-and-fire neurons, exact reductions and synchrony-aware rate equations show why retaining subthreshold state variables can be essential for macroscopic dynamics \cite{MontbrioEtAl2015,devalle2017synchrony}. The finite-jump generator preserves shot-noise variability associated with a specified event intensity; the recurrent closure, by contrast, replaces the cell-conditioned statistics of that intensity by a global mean.

The difference between the Type I and II estimators is especially transparent after defining $r_t^{XY}=\bar H_t^{XY}/\tau^{XY}$, the mean rate at which pending type-$Y$ inputs act on population $X$. Equation~\eqref{eq:mean_drive} becomes
\begin{equation}
 \tau^{XY}\frac{d r_t^{XY}}{dt}=-r_t^{XY}+c^{XY}f_t^Y.
 \label{eq:discussion_filter}
\end{equation}
When using the Type I estimator, $r^{XY}=c^{XY}f^Y$, and the synaptic timescale disappears. Hence, a stationary balance cannot represent rate changes produced by the attenuation and phase lag of this filter. The Type II estimator retains the filter and averages the coupled nonlinear flux. In general,
\begin{equation}
 \left\langle\bm\rho_t^XK_X(\bar{\bm H}_t^X)\mathbf{1}\right\rangle_t
 \neq
 \left\langle\bm\rho_t^X\right\rangle_t
 K_X\!\left(\left\langle\bar{\bm H}_t^X\right\rangle_t\right)\mathbf{1}.
 \label{eq:discussion_covariation}
\end{equation}
The Type II estimator thus preserves the joint temporal evolution of the population occupancy and mean drive. It does not preserve the instantaneous neuron-level joint law of state and pending input or microscopic pairwise spike correlations.

This construction builds on a sequence of Markov reductions of spiking networks. Earlier work retained coarse population counts and aggregate pending-input pools needed to describe gamma-band dynamics \cite{cai2021model}; related population-distribution and conductance coordinates were used to study multi-band oscillations of spiking neuronal networks \cite{wu2023multi}. The more recent dsODE framework evolves a discretized voltage occupancy and the mean and variance of pending recurrent inputs; its stochastic companion also represents finite-population fluctuations \cite{chang2025minimizing}. At the level of resolved variables, the Type II estimator is a first-moment analogue of that framework: it retains occupancy and a global mean pending-input pool, omits pending-pool variance, and extends the state partition and spike-detection construction to two- and three-dimensional intrinsic dynamics. The Type I estimator further ignores the temporal dynamics of recurrent input and obtains a stationary closure.


Compared to other classical population methods: A Fokker--Planck equation may exactly describe the evolution of voltage distribution, yet the diffusion approximation can distort firing rates when individual synaptic jumps are not small \cite{OmurtagEtAl2000,richardson2010shotnoise}. This is exactly pronounced during partial synchrony in the dynamics, which can produce rate bias missed by mean-field reductions \cite{ocker2017nonlinear,li2019well}. Field-theoretic and system-size expansions make the feedback of correlations and higher cumulants explicit \cite{buice2010systematic,buice2013dynamic}, while mesoscopic equations retain finite-population activity fluctuations \cite{schwalger2017towards}. Our work belongs to this family, with further reductions to focus on the estimation of mean firing rates. 

A natural next step is to replace the global first-moment closure by a state-conditioned hierarchy. For each phase-space cell, one could first evolve the conditional mean and variance of each pending-input pool. A richer closure could then add excitatory–inhibitory cross-covariances and selected covariances between voltage and recurrent drive. These quantities would distinguish an input fluctuation delivered near the firing surface from the same fluctuation delivered far from it. Kinetic, maximum-entropy, and conductance-based closures, together with the global input-moment equations in dsODE, provide useful starting points \cite{cai2006kinetic,pmid16712338,rangan2008conductance,chang2025minimizing}. Sparse or adaptive partitions could control the added cost, yielding a principled sequence of reductions that connects phase-space resolution and input statistics to firing-rate accuracy.

\begin{acknowledgments}
This work was partially supported by the National Natural Science and Technology Innovation STI2030-Major Project No.~2022ZD0204600 and the Natural Science Foundation of China through Grant No.~31771147 (Z.W. and L.T.). Z.-C.X. was supported by the Courant Institute of Mathematical Sciences at New York University during most of this study and subsequently by the Shanghai Municipal Education Commission through the Eastern Talent Program. We thank Aaditya Rangan (New York University) for helpful discussions.
\end{acknowledgments}

\section*{Author Declarations}

\subsection*{Conflict of Interest}

The authors declare no conflict of interest.

\subsection*{Ethics Approval}

Ethics approval is not required for this computational study.

\subsection*{Author Contributions}

Conceptualization: Z.-C.X., L.T., and Z.W.; Methodology: Z.W., Z.-C.X., and L.T.; Software: Z.W.; Data curation: Z.W.; Writing---original draft: Z.-C.X., L.T., and Z.W.; Writing---review and editing: Z.-C.X., L.T., and Z.W.; Supervision: Z.-C.X.

\section*{Data Availability}

The simulation outputs, parameter sets, and analysis code supporting this study are available at \url{https://github.com/ZhongyiWang01/MarkovNeuronalNetwork}.

\appendix

\section{Model-specific equations and spike conventions}
\label{app:model_details}

\subsection{LIF and EIF models}

For an LIF neuron, $\bm z_i^X=V_i^X$ and
\begin{equation}
 \frac{dV_i^X}{dt}=g^{X,\mathrm{leak}}(V^{\mathrm{rest}}-V_i^X)
 +J_i^X(t,V_i^X).
 \label{eq:lif_model}
\end{equation}
When $V_i^X$ reaches $v^{X,\mathrm{th}}$, the neuron emits a spike and enters a refractory state $\mathcal R_X$ for a period $\tau_{\mathrm{ref}}^X$, after which it is reset to $V^{\mathrm{rest}}$. Leakage constants are taken as $g^{E,\mathrm{leak}} = g^{I,\mathrm{leak}} = (20\,\rm{ms})^{-1}$.
The direct model keeps the neuron unresponsive during this period.

The EIF model retains the same state and threshold--refractory--reset rule but uses
\begin{align}
 \frac{dV_i^X}{dt}&=g^{X,\mathrm{leak}}
 \left[V^{\mathrm{rest}}-V_i^X+\Delta_T^X
 \exp\left(\frac{V_i^X-V_T^X}{\Delta_T^X}\right)\right] \nonumber\\
 &+J_i^X(t,V_i^X).
 \label{eq:eif_model}
\end{align}
Here $V_T^X$ controls the onset of exponential spike initiation, whereas $\Delta_T^X$ controls its sharpness. In Eq.~\eqref{eq:eif_model}, the exponential term is inside the leakage prefactor, and $J_i^X$ is additive \cite{Fourcaud-Trocme2003-lm}.

\subsection{FHN model}

For the FHN model \cite{fitzhugh1961impulses}, $\bm z_i^X=(v_i^X,w_i^X)$ and
\begin{align}
 \frac{dv_i^X}{dt}&=v_i^X-\frac{(v_i^X)^3}{3}-w_i^X
 +J_{i}^X(t,v_i^X),
 \label{eq:fhn_v}\\
 \frac{dw_i^X}{dt}&=\varepsilon^X(v_i^X+a^X-b^Xw_i^X),
 \label{eq:fhn_w}
\end{align}
where $\varepsilon^X\ll1$ separates the fast and slow timescales, and $a^X$, $b^X$, and the external input current $I_0^X$ in $J^X$ set the excitable or oscillatory regime. A spike is recorded at an upward crossing of $v^{X,\mathrm{th}}$, with no refractory state or reset. In this formulation, detected spikes are readouts and do not alter the network dynamics. 

\subsection{Two-dimensional reduced HH model}
In both HH models, $V$ is measured in mV and the resulting transition rates are in ms$^{-1}$; consequently, $\tau_w$ is measured in ms.
The HH-2D state is $\bm z_i^X=(V_i^X,w_i^X)$, with
\begin{align}
 C^X\frac{dV_i^X}{dt}
 &=J_i^X(t,V_i^X)-\bar g_L^X(V_i^X-E_L)
 \nonumber\\
 &\quad-\bar g_{\mathrm{Na}}^Xm_\infty^3(V_i^X)h(w_i^X)
 (V_i^X-E_{\mathrm{Na}})
 \nonumber\\
 &\quad-\bar g_{\mathrm{K}}^Xn^4(w_i^X)(V_i^X-E_{\mathrm{K}}),
 \label{eq:hh2d_v}\\
 \frac{dw_i^X}{dt}
 &=\frac{w_\infty(V_i^X)-w_i^X}{\tau_w(V_i^X)}.
 \label{eq:hh2d_w}
\end{align}
Here $C^X$ is the membrane capacitance; $\bar g_L^X$, $\bar g_{\mathrm{Na}}^X$, and $\bar g_{\mathrm{K}}^X$ are maximal conductances; and $E_L$, $E_{\mathrm{Na}}$, and $E_{\mathrm{K}}$ are reversal potentials. An upward crossing of $v^{X,\mathrm{th}}$ is counted as a spike and generates synaptic input without resetting the state \cite{hodgkin1952quantitative, rinzel1985excitation}.

The relevant gating functions are given by

\begin{align}
    h(w)&=h_0-w,\qquad h_0=0.8,\\
n(w)&=w,\\
m_\infty(V)&=\frac{\alpha_m(V)}{\alpha_m(V)+\beta_m(V)},\label{eq:hh_m_infty}\\
w_\infty(V)&=n_\infty(V)=\frac{\alpha_n(V)}{\alpha_n(V)+\beta_n(V)},\\
\tau_w(V)&=\frac{1}{\alpha_n(V)+\beta_n(V)},\\
\alpha_m(V)&=\frac{0.1(V+40)}{1-\exp(-(V+40)/10)}, \\
\beta_m(V)&=4\exp(-(V+65)/18), \\
\alpha_n(V)&=\frac{0.01(V+55)}{1-\exp(-(V+55)/10)}, \\
\beta_n(V)&=0.125\exp(-(V+65)/80). \label{eq:hh_beta_n}
\end{align}

\subsection{Three-dimensional reduced HH model}

The HH-3D state is $\bm z_i^X=(V_i^X,n_i^X,h_i^X)$. Sodium activation is replaced by its voltage-dependent steady-state value $m_\infty(V)$, giving
\begin{align}
 C^X\frac{dV_i^X}{dt}
 &=J_i^X(t,V_i^X)-\bar g_L^X(V_i^X-E_L)
 \nonumber\\
 &\quad-\bar g_{\mathrm{Na}}^Xm_\infty^3(V_i^X)h_i^X
 (V_i^X-E_{\mathrm{Na}})
 \nonumber\\
 &\quad-\bar g_{\mathrm{K}}^X(n_i^X)^4(V_i^X-E_{\mathrm{K}}),
 \label{eq:hh3d_v}\\
 \frac{dn_i^X}{dt}
 &=\alpha_n(V_i^X)(1-n_i^X)-\beta_n(V_i^X)n_i^X,
 \label{eq:hh3d_n}\\
 \frac{dh_i^X}{dt}
 &=\alpha_h(V_i^X)(1-h_i^X)-\beta_h(V_i^X)h_i^X.
 \label{eq:hh3d_h}
\end{align}
An upward crossing of $v^{X,\mathrm{th}}$ is counted as a spike and generates synaptic input without resetting the state. The conductance structure is based on the HH model \cite{hodgkin1952quantitative}, with sodium activation replaced by its voltage-dependent steady-state value \cite{meunier1992dimreductionHH}.

The voltage-dependent transition rates for $h_i^X$ are
\begin{align}
    \alpha_h(V)&=0.07\exp(-(V+65)/20),\\
    \beta_h(V)&=\frac{1}{1+\exp(-(V+35)/10)}.
\end{align}
$m_\infty(V), \alpha_n(V)$, and $\beta_n(V)$ are the same as in the two-dimensional case in Eqs. \eqref{eq:hh_m_infty}--\eqref{eq:hh_beta_n}. 

\section{Multidimensional transition construction and firing flux}
\label{app:discretization}

The construction separates continuous phase-space transport from inputs that instantaneously change a neuron's intrinsic state. Let
$\bm B_X(\bm z;\bm\theta_t)$ denote the total continuous vector field for
population $X$, where $\bm\theta_t$ collects any population quantities on
which that field depends. Thus, $\bm B_X$ includes the intrinsic vector
field for single cells, the external constant input, and any recurrent or external spiking input. If $\Phi_X^{\Delta t}$ is its short-time flow
map and $\Pi_X^{\mathrm{flow}}(\Delta t;\bm\theta_t)$ is the corresponding
row-stochastic redistribution of cell mass, the conservative drift
generator can be written formally as
\begin{equation}
 D_X(\bm\theta_t)
 =\lim_{\Delta t\downarrow0}
 \frac{\Pi_X^{\mathrm{flow}}(\Delta t;\bm\theta_t)-\mathsf I_X}
 {\Delta t}.
 \label{eq:drift_generator}
\end{equation}
Below, $\mathcal D_X[\bm B_X]$ denotes this conservative
discretization of the specified vector field.
An equivalent finite-volume or local upwind construction can be used. In
each case, the off-diagonal entries of $D_X$ are nonnegative, and the
diagonal entries are chosen so that $D_X\mathbf{1}=\bm{0}$ after the stated
boundary and firing rules have been applied.

For an input that acts as an instantaneous event, let
$\mathcal T_X^\eta$ be its model-specific state map, where
$\eta\in\{\mathrm{ext},XE,XI\}$. If $\Pi_X^\eta$ is a row-stochastic
redistribution matrix whose entry $[\Pi_X^\eta]_{\alpha\beta}$ gives the
interpolation or cell-average weight transferred from $C_\alpha^X$ to
$C_\beta^X$ under $\mathcal T_X^\eta$, then the corresponding unit-event
generator is
\begin{equation}
 J_X^\eta=\Pi_X^\eta-\mathsf I_X.
 \label{eq:event_generator}
\end{equation}
Multiplying $J_X^\eta$ by the corresponding event rate gives its
contribution to $Q_X$. A constant current is therefore included in
$D_X$, whereas an instantaneous Poisson kick contributes
$\lambda^{X,\mathrm{ext}}J_X^{\mathrm{ext}}$. A filtered input instead
requires a pending-input or synaptic-activation variable and is not, in
general, represented by the instantaneous-event term alone.
Under the assumed within-population homogeneity, we set $I_i^0\equiv I_0^X$. If it varied across neurons, a
single population generator should be replaced by bias-conditioned
generators or by an explicitly stated heterogeneity closure.
In the model-specific total fields below, $\bm U_X$ denotes the contribution of intrinsic single-cell dynamics, recurrent input and external Poissonian kicks, so that the external constant input
$I_0^X$ is separated from other current terms.

Within this construction, LIF and EIF drift or input transitions that
reach or cross the firing threshold are redirected to $\mathcal R_X$,
and $R_X$ transfers mass from $\mathcal R_X$ to the reset cell. For FHN
and HH models, the state is not redirected: a directed transition from
below to above the detection surface is tagged as a member of
$\mathcal E_X^{\mathrm{fire}}$ while otherwise following the ordinary
phase-space transport. This edge-based rule assumes that the detection
surface is aligned with cell faces; if it cuts through a cell, the drift
discretization must instead retain the corresponding subcell boundary
flux. These tagging rules yield Eq.~\eqref{eq:firing_flux} without
requiring a refractory state.

\subsection{One-dimensional LIF specialization}

The reported LIF implementation uses a uniform voltage grid
\begin{equation}
 \Gamma_X=\{-m^I,-m^I+1,\ldots,m^{X,\mathrm{th}}-1\},
 \quad
 \widehat\Gamma_X=\Gamma_X\cup\{\mathcal R_X\},
 \label{eq:lif_grid}
\end{equation}
where bin $m$ represents $[ma,(m+1)a)$ and
\begin{equation}
 a=\frac{v^{X,\mathrm{th}}-V^I}{m^{X,\mathrm{th}}+m^I}.
 \label{eq:lif_bin_width}
\end{equation}
For a dimensionless implementation of an LIF neuron with inhibitory reversal potential $V^I=-2/3$, $V_\mathrm{rest} = 0$ and $V^{\rm th} = 1$, state $-m^I$ represents bin $[-2/3, -2/3+a)$ and $m^{X,\mathrm{th}}-1$ corresponds to bin $[1-a,1)$.

To distinguish physical input strength from displacement on the voltage
grid, let $\sigma^{XY}=\frac{S^{XY}}{a}$ and $\sigma^{X,\mathrm{ext}}=\frac{S^{X,\mathrm{ext}}}{a}$ denote the
dimensionless jump magnitudes, in bin units, produced by recurrent and
external events, respectively. For a noninteger jump, the
scheme splits the transferred mass between the two neighboring
destination bins. Writing $\lfloor x\rfloor$ for the floor and
$\{x\}=x-\lfloor x\rfloor$ for the fractional part, a positive jump of
size $\sigma$ sends mass to $m+\lfloor\sigma\rfloor$ and
$m+\lfloor\sigma\rfloor+1$ with weights $1-\{\sigma\}$ and
$\{\sigma\}$, respectively. Destinations at or above threshold are
redirected to $\mathcal R_X$. The inhibitory jump size is voltage
dependent,
\begin{equation}
 \widetilde\sigma^{XI}(m)=\sigma^{XI}
 \frac{m+m^I}{m^{X,\mathrm{th}}+m^I},
 \label{eq:inhibitory_jump}
\end{equation}
and is directed toward lower voltages. More generally, the LIF drift
matrix is constructed from the full continuous voltage velocity,
\begin{equation}
 B_X^{\mathrm{LIF}}(V)=U_{X,V}^{\mathrm{LIF}}(V)+I_0^X,
 \qquad
 D_X^{\mathrm{LIF}}(I_0^X)=\mathcal D_X[B_X^{\mathrm{LIF}}].
 \label{eq:lif_total_drift}
\end{equation}

For $I_0^X=0$ under the reset-at-zero scaling, the reported
nearest-neighbor leakage rule moves the state one bin toward reset at
rate $|m|g^{X,\mathrm{leak}}$; the absolute value ensures a nonnegative
transition rate. When $I_0^X\ne0$,
the direction and rate of every subthreshold drift transition must instead
be obtained from Eq.~\eqref{eq:lif_total_drift}. The refractory state
returns to the reset bin at rate $1/\tau_{\mathrm{ref}}^X$.

For an upward jump of size $\sigma$, define the probability mass transferred across threshold by the following expression, with the convention
$\rho_{t,m}^X=0$ for $m\notin\Gamma_X$:
\begin{equation}
 \mathcal M_t^X(\sigma)
 =\sum_{\substack{m\in\Gamma_X\\
 m\ge m^{X,\mathrm{th}}-\lfloor\sigma\rfloor}}
 \rho_{t,m}^X
 +\{\sigma\}\rho_{t,m^{X,\mathrm{th}}-\lfloor\sigma\rfloor-1}^X.
 \label{eq:lif_crossing_mass}
\end{equation}
The excitatory recurrent and external contributions to the firing flux
are then
\begin{equation}
 f_t^X
 =\frac{\bar H_t^{XE}}{\tau^{XE}}
 \mathcal M_t^X(\sigma^{XE})
 +\lambda^{X,\mathrm{ext}}
 \mathcal M_t^X(\sigma^{X,\mathrm{ext}}).
 \label{eq:lif_explicit_flux}
\end{equation}
Equation~\eqref{eq:lif_explicit_flux} is the LIF instance of the tagged
flux in Eq.~\eqref{eq:firing_flux} when the continuous drift has no
outward threshold flux. Under that condition, its stationary value
agrees with Eq.~\eqref{eq:refractory_flux}; the threshold index itself is
not a state in Eq.~\eqref{eq:lif_grid}.

\subsection{One-dimensional EIF specialization}

The EIF construction uses the LIF voltage
partition, refractory state, input-event redistribution matrices, and
reset operator, but replaces the linear subthreshold drift by the EIF velocity. With the homogeneous constant current denoted by $I_0^X$, let
\begin{equation}
 B_X^{\mathrm{EIF}}(V)
 =U_{X,V}^{\mathrm{EIF}}(V)+I_0^X
 \label{eq:eif_total_velocity}
\end{equation}
be the full continuous voltage drift specified by the EIF equation.
Discretizing this velocity according to Eq.~\eqref{eq:drift_generator}
defines $D_X^{\mathrm{EIF}}(I_0^X)$. The constant current is therefore
part of the drift matrix rather than a separate jump term.

Because LIF and EIF use the same input architecture, their external and recurrent event matrices are constructed with the same voltage-jump and interpolation rules. The two models only differ in the continuous drift operator. Any drift or input transition that reaches or crosses
$v^{X,\mathrm{th}}$ is sent to $\mathcal R_X$, and $R_X$ returns
refractory mass to the reset cell. Drift and input events leave
$\mathcal R_X$ unchanged, so $R_X$ is its only outflow. Under the
instantaneous-external-input convention,
\begin{align}
 Q_X^{\mathrm{EIF}}(\bar{\bm H}_t^X)
 &=D_X^{\mathrm{EIF}}(I_0^X)
 +\lambda^{X,\mathrm{ext}}J_X^{\mathrm{ext}} \nonumber
 \\
 &+\sum_{Y\in\{E,I\}}
 \frac{\bar H_t^{XY}}{\tau^{XY}}J_X^{XY}
 +R_X.
 \label{eq:eif_generator}
\end{align}
Unlike passive LIF leakage, the exponential EIF drift can itself carry
mass across threshold. The EIF firing flux must therefore retain the
drift contribution as well as the external and recurrent event
contributions:
\begin{align}
 f_t^X=\sum_{m\in\Gamma_X}\rho_{t,m}^X(&[D_X^{\mathrm{EIF}}]_{m,\mathcal R_X}+\lambda^{X,\mathrm{ext}}[J_X^{\mathrm{ext}}]_{m,\mathcal R_X} \nonumber \\
 &+\sum_{Y\in\{E,I\}}\frac{\bar H_t^{XY}}{\tau^{XY}}[J_X^{XY}]_{m,\mathcal R_X}).
 \label{eq:eif_firing_flux}
\end{align}
At stationarity, Eq.~\eqref{eq:eif_firing_flux} is equivalently given by
$f_\infty^X=\rho_{\infty,\mathcal R_X}^X/\tau_{\mathrm{ref}}^X$.

\subsection{Two-dimensional FHN specialization}

The FHN state cells partition the $(v,w)$ plane. In the direct network,
define the empirical voltage mean by
\begin{equation}
 \bar v_X^{\mathrm{emp}}(t)
 =\frac{1}{N_X}\sum_{i\in X}v_i^X(t).
 \label{eq:fhn_empirical_mean_voltage}
\end{equation}
For the same-population term, exclusion of the target neuron then gives
the exact identity
\begin{equation}
 \sum_{\substack{j\in X\\j\ne i}}(v_j^X-v_i^X)
 =N_X\left(\bar v_X^{\mathrm{emp}}-v_i^X\right),
 \label{eq:fhn_empirical_gap_identity}
\end{equation}
because the omitted self term is zero.

Let $(v_\alpha^\ast,w_\alpha^\ast)$ be the representative state of
$C_\alpha^X$. The reduced occupancy approximates the population mean by
\begin{equation}
 \bar v_X^{\mathrm{red}}(t)=\sum_{\alpha\in\Gamma_X}
 \rho_{t,\alpha}^Xv_\alpha^\ast.
 \label{eq:fhn_reduced_mean_voltage}
\end{equation}
When $\bm\rho_t^X$ represents the empirical occupancy, replacing the
within-cell voltages by $v_\alpha^\ast$ turns the exact all-to-all
electrical-coupling current into
\begin{align}
 \mathcal G_X(v;\bar{\bm v}_t^{\mathrm{red}})
 &=\sum_{Y\in\{E,I\}}N_YG_{\mathrm{gap}}^{XY}
 \left[\bar v_Y^{\mathrm{red}}(t)-v\right], \nonumber \\
 \bar{\bm v}_t^{\mathrm{red}}
 &=(\bar v_E^{\mathrm{red}}(t),\bar v_I^{\mathrm{red}}(t)).
 \label{eq:fhn_reduced_gap_current}
\end{align}
If $\bm\rho_t^X$ is instead a representative-neuron marginal, an
exchangeable mean-field closure replaces the conditional mean voltages of
the other neurons by the unconditional population means. Under that
closure, the same-population term carries $N_X-1$ rather than $N_X$; the
choice must follow the interpretation of $\bm\rho_t^X$ in the main text.

The full continuous FHN vector field used to construct the drift matrix
is
\begin{equation}
 \bm B_X^{\mathrm{FHN}}(v,w;\bar{\bm v}_t^{\mathrm{red}})
 =\bm F_X^{\mathrm{FHN}}(v,w)
 +\bm e_V
 \left[I_0^X+
 \mathcal G_X(v;\bar{\bm v}_t^{\mathrm{red}})\right].
 \label{eq:fhn_total_drift}
\end{equation}
The constant and recurrent conductance currents therefore enter the continuous transport operator,
\begin{equation}
 D_X^{\mathrm{FHN}}(\bar{\bm v}_t^{\mathrm{red}};I_0^X)
 =\mathcal D_X\!\left[
 \bm B_X^{\mathrm{FHN}}(\,\cdot\,;\bar{\bm v}_t^{\mathrm{red}})
 \right],
 \label{eq:fhn_drift_matrix}
\end{equation}
and neither is represented by a recurrent spike-event matrix.

For instantaneous external input, one Poisson event changes only the
voltage-like coordinate,
\begin{equation}
 \mathcal T_X^{\mathrm{ext}}(v,w)
 =\left(v+S^{X,\mathrm{ext}},w\right),
 \label{eq:fhn_external_map}
\end{equation}
where $S^{X,\mathrm{ext}}$ is the time integral of the external current
impulse and, because the FHN voltage equation has unit capacitance, is
also the resulting voltage increment. Redistribution under this map gives
$J_X^{\mathrm{ext}}$, and the FHN generator is
\begin{equation}
 Q_X^{\mathrm{FHN}}(\bar{\bm v}_t^{\mathrm{red}})
 =D_X^{\mathrm{FHN}}(\bar{\bm v}_t^{\mathrm{red}};I_0^X)
 +\lambda^{X,\mathrm{ext}}J_X^{\mathrm{ext}}.
 \label{eq:fhn_generator}
\end{equation}
There are no recurrent event matrices and no refractory-release
operator: gap junctions act continuously, and a detected FHN spike is a
readout rather than a recurrent event. Because the means in
Eq.~\eqref{eq:fhn_reduced_mean_voltage} depend on the current
occupancies, the two population forward equations are coupled and
nonlinear. A Type II estimator updates $\bar v_E^{\mathrm{red}}(t)$ and
$\bar v_I^{\mathrm{red}}(t)$ from the current cell probabilities before constructing
the drift matrices; a stationary calculation requires the corresponding
self-consistency.

An upward transition across the prescribed FHN detection surface is
included in $\mathcal E_X^{\mathrm{fire}}$ but retains its ordinary
destination cell. The firing rate follows from Eq.~\eqref{eq:firing_flux},
with no redirection, reset, or feedback through a recurrent spike pool.

\subsection{Two- and three-dimensional HH specializations}

The HH-2D and HH-3D generators have the same operator structure and
differ only in their intrinsic drift and gating coordinates. HH-2D uses
cells in $(V,w)$, whereas HH-3D uses cells in $(V,n,h)$. In both cases,
the synaptic activation and conductance parameters enter at distinct
stages. For a connection from population $Y$ to population $X$,
$a^{XY}$ is the increment of the synaptic activation variable caused by
one arriving spike. We write
$g^{XY}\equiv\bar g_{ij}^{\mathrm{syn}}$ for the
population-homogeneous conductance per unit activation.
Likewise,
$E^{XY}$ is the corresponding synaptic reversal
potential, denoted by $V^Y$ in the common notation of
Sec.~\ref{sec:models}.

For each delivered connection $j\to i$, let $s_{ij}(t)$ be its dimensionless synaptic activation. Between arrivals it obeys $ds_{ij}/dt=-s_{ij}/\tau^{XY}$, and each delivered spike increments it by $a^{XY}$; for absent connections, $s_{ij}\equiv0$.
\begin{equation}
 \bar s_t^{XY}
 =\mathbb E\!\left[
 \sum_{\substack{j\in Y\\(Y,j)\ne(X,i)}}s_{ij}(t)
 \right].
 \label{eq:hh_mean_activation_definition}
\end{equation}
Hence $\bar s_t^{XY}$ is the expected sum over incoming contacts, not a
per-synapse mean. Its first-moment equation is
\begin{equation}
 \frac{d\bar s_t^{XY}}{dt}
 =-\frac{\bar s_t^{XY}}{\tau^{XY}}
 +a^{XY}r_{\mathrm{arr}}^{XY}(t),
 \label{eq:hh_mean_activation}
\end{equation}
where $r_{\mathrm{arr}}^{XY}(t)$ is the mean rate at which delivered
type-$Y$ spikes arrive at one type-$X$ neuron and $c^{XY}$ is the
connectivity factor that converts the per-neuron population rate $f^Y$
to that arrival rate, hence $r_{\mathrm{arr}}^{XY}(t)=c^{XY}f_t^Y$. With independent delivery probability $P^{XY}$ and
excluded autopases, $c^{XY}=N_YP^{XY}$ for $Y\ne X$ and
$c^{XX}=(N_X-1)P^{XX}$.

With the normalized exponential kernel in
Eq.~\eqref{eq:synaptic_kernel}, a dimensionless activation variable and
homogeneous synapses give
\begin{equation}
 S^{XY}=\tau^{XY}a^{XY}g^{XY}.
 \label{eq:hh_common_strength}
\end{equation}
Here $S^{XY}$ is the time-integrated conductance strength in the common
kernel notation. If $\bar H_t^{XY}$ satisfies the zero-delay pending-pool
equation with matched initial data, then
\begin{equation}
 \bar s_t^{XY}=a^{XY}\bar H_t^{XY},
 \qquad
 g^{XY}\bar s_t^{XY}
 =\frac{S^{XY}}{\tau^{XY}}\bar H_t^{XY}.
 \label{eq:hh_pending_pool_crosswalk}
\end{equation}
This identity provides the crosswalk to a formulation parameterized by
$\bar H_t^{XY}$ while leaving the conductance current in the continuous
drift. Under the supplied linear activation and current equations,
$a^{XY}$ and $g^{XY}$ affect voltage through their product
for zero or correspondingly scaled activation initial data; their
separate symbols describe different mechanistic stages rather than
independent coupling combinations.

The mean-activation closure also invokes the factorization
\begin{equation}
 \mathbb E\!\left[
 \left.
 \sum_{\substack{j\in Y\\(Y,j)\ne(X,i)}}s_{ij}(t)
 \right|\bm z_i^X(t)=\bm z
 \right]
 \approx \bar s_t^{XY}.
 \label{eq:hh_activation_factorization}
\end{equation}
Neglecting these correlations permits a single intrinsic-state generator for all cells. Write
$\bar{\bm s}_t^X=(\bar s_t^{XE},\bar s_t^{XI})$. Under this closure, the
total continuous vector field for the $q$-dimensional HH reduction,
$q\in\{2,3\}$, is
\begin{align}
 \bm B_{X,q}^{\mathrm{HH}}(\bm z;\bar{\bm s}_t^X)
 =&~\bm F_{X,q}^{\mathrm{HH}}(\bm z)\nonumber \\
 &+\frac{\bm e_V}{C^X}
 \left[
 I_0^X+
 \sum_{Y\in\{E,I\}}
 g^{XY}\bar s_t^{XY}
 \left(E^{XY}-V\right)
 \right].
 \label{eq:hh_effective_drift}
\end{align}
Here $\bm F_{X,2}^{\mathrm{HH}}$ contains the intrinsic voltage and
$w$ dynamics, while $\bm F_{X,3}^{\mathrm{HH}}$ contains the intrinsic
voltage, $n$, and $h$ dynamics. Thus, HH-2D transport crosses cell faces
in the $V$ and $w$ directions, whereas HH-3D transport can additionally
cross faces in the $n$ and $h$ directions. Discretizing all components of
Eq.~\eqref{eq:hh_effective_drift} defines
\begin{equation}
 D_{X,q}^{\mathrm{HH}}(\bar{\bm s}_t^X;I_0^X)
 =\mathcal D_X\!\left[
 \bm B_{X,q}^{\mathrm{HH}}(\,\cdot\,;\bar{\bm s}_t^X)
 \right].
 \label{eq:hh_drift_matrix}
\end{equation}
Accordingly, the constant current and the recurrent conductance current
belong to continuous phase-space transport.

If the external Poisson input is instantaneous and additive, its event
map is
\begin{equation}
 \mathcal T_{X,q}^{\mathrm{ext}}(V,\bm u)
 =\left(V+\frac{S^{X,\mathrm{ext}}}{C^X},\bm u\right),
 \label{eq:hh_external_map}
\end{equation}
where $S^{X,\mathrm{ext}}$ is the time integral of the external current
impulse, $\bm u=(w)$ for HH-2D, and $\bm u=(n,h)$ for HH-3D. The gating
coordinates are unchanged by the instantaneous event. The resulting
intrinsic-state generator is
\begin{equation}
 Q_{X,q}^{\mathrm{HH}}(\bar{\bm s}_t^X)
 =D_{X,q}^{\mathrm{HH}}(\bar{\bm s}_t^X;I_0^X)
 +\lambda^{X,\mathrm{ext}}J_{X,q}^{\mathrm{ext}},
 \qquad q\in\{2,3\}.
 \label{eq:hh_generator}
\end{equation}
Equation~\eqref{eq:hh_generator} is the conditional generator of the
intrinsic HH coordinates. The full reduced dynamics couple
$d\bm\rho_t^X/dt=\bm\rho_t^XQ_{X,q}^{\mathrm{HH}}
(\bar{\bm s}_t^X)$ to Eq.~\eqref{eq:hh_mean_activation}. There is no
refractory-release operator. In this mean-activation formulation there
is also no recurrent jump matrix acting directly on $(V,w)$ or
$(V,n,h)$: a recurrent arrival changes the activation in
Eq.~\eqref{eq:hh_mean_activation}, and the activation changes voltage
through Eq.~\eqref{eq:hh_effective_drift}. A first-order exponentially
filtered external input can be treated analogously by evolving its mean
activation and including the associated current in the continuous drift;
other filter dynamics require their own pending-input or augmented-state
description.

An upward transition across $V=v^{X,\mathrm{th}}$ is tagged as a firing
transition but retains its ordinary destination cell. Hence, spike
detection does not reset the HH state. Unlike FHN, however, the resulting
firing flux enters the arrival rates in Eq.~\eqref{eq:hh_mean_activation}
and thereby provides recurrent feedback.

\section{Numerical solution and diagnostics}
\label{app:numerics}

The Type II estimator uses the forward-Euler method, with the default settings summarized in Table~\ref{tab:type2_settings}. At each step, the explicit scheme evaluates $Q_X$ and $f_n^X$ from $(\bm\rho_n,\bar{\bm H}_n)$ and then applies
\begin{align}
 \bm\rho_{n+1}^X
 &=\bm\rho_n^X+\Delta t\,\bm\rho_n^XQ_X(\bar{\bm H}_n^X),
 \label{eq:euler_rho}\\
 \bar H_{n+1}^{XY}
 &=\bar H_n^{XY}+\Delta t
 \left(-\frac{\bar H_n^{XY}}{\tau^{XY}}+c^{XY}f_n^Y\right).
 \label{eq:euler_h}
\end{align}
The running firing-rate average can then be updated from the same $f_n^X$. The scheme initializes $\bar H_0^{XY}=0$ and the running firing-rate average at zero. The stated stopping rule requires the changes in the excitatory and inhibitory running firing-rate averages to remain below $\varepsilon$ for $K$ consecutive steps, with a maximum of $N_{\max}$ steps.


\begin{table}[htbp]
\caption{Default numerical settings for the forward-Euler scheme in the Type II estimator. The stated point-mass initialization applies to the one-dimensional implementation.}
\label{tab:type2_settings}
\begin{ruledtabular}
\begin{tabular}{lcc}
Quantity & Meaning & Default \\
\hline
$\varepsilon$ & Convergence tolerance & $10^{-3}$ \\
$K$ & Consecutive accepted steps & $5\times10^3$ \\
$N_{\max}$ & Maximum number of steps & $3\times10^5$ \\
$\Delta t$ & Step size (s) & $10^{-5}$ \\
$\bm\rho_0^X$ & Initial LIF distribution & $\rho_0^X(v)=\mathds{1}_{v=0}$ \\
\end{tabular}
\end{ruledtabular}
\end{table}

All simulations and reduced numerical solutions were performed in MATLAB R2024b (The MathWorks, Inc.) with the Parallel Computing Toolbox on a Windows 11 desktop with an Intel Core i9-13900K processor and 64~GB RAM. No GPU acceleration was used.

\subsection{Simulation-window estimate for network firing rates}
\label{app:rate-sampling-window}
Here we explain our choice of network simulation time $T = 10$ to $100$~s.

For a stationary homogeneous population of \(N\) neurons, let \(K_i(T)\) denote the spike count of neuron \(i\) over a post-transient measurement window of duration \(T\), and let \(f\) be the mean firing rate per neuron, so that \(\mathbb{E}[K_i(T)]=fT\). We denote by \(F_1=N^{-1}\sum_i\operatorname{Var}[K_i(T)]/\mathbb{E}[K_i(T)]\) the neuronwise-averaged Fano factor and by \(\rho\) the average pairwise correlation of the counts \(K_i(T)\) over the same window. Under the exchangeability approximation, \(\operatorname{Var}[K_i(T)]=F_1fT\) and \(\operatorname{Cov}[K_i(T),K_j(T)]=\rho F_1fT\) for \(i\neq j\). The estimator of the population-averaged firing rate,
\begin{equation}
    \widehat f_T=\frac{1}{NT}\sum_{i=1}^{N}K_i(T),
\end{equation}
therefore satisfies
\begin{align}
    \frac{\sqrt{\operatorname{Var}(\widehat f_T)}}{f}
    &=
    \sqrt{\frac{F_1[1+(N-1)\rho]}{NfT}}.
    \label{eq:network-rate-sampling-error}
\end{align}
Thus, synchrony reduces the effective number of independent neurons from \(N\) to \(N_{\mathrm{eff}}=N/[1+(N-1)\rho]\). Here \(\rho\) is a count correlation over the full estimation window, rather than the short-bin synchrony index used elsewhere in this work; the same calculation can be applied separately to the excitatory and inhibitory populations using their population-specific parameters.

Now we estimate the model time required to constrain the relative uncertainty of the firing-rate estimate. Under a central-limit approximation, requiring a two-sided \(1-\alpha\) confidence interval for \(f\) to have relative half-width at most \(\epsilon\) gives
\begin{equation}
    T
    \gtrsim
    \frac{z_{1-\alpha/2}^{\,2}F_1}{\epsilon^2f}
    \left(\rho+\frac{1-\rho}{N}\right).
    \label{eq:required-rate-window}
\end{equation}
For a \(95\%\) interval and \(\epsilon=0.2\), this becomes
\begin{equation}
    T_{95,20\%}
    \gtrsim
    \frac{96.04F_1}{f}
    \left(\rho+\frac{1-\rho}{N}\right).
    \label{eq:required-rate-window-20}
\end{equation}
Once \(N\rho\gg1\), the required duration is nearly independent of network size and scales as \(T_{95,20\%}\simeq96.04F_1\rho/f\). For example, at a firing rate \(f=10\,\mathrm{Hz}\), the conservative choices \(F_1=10\) and \(\rho=0.5\) give \(T_{95,20\%}\simeq50\,\mathrm{s}\) for \(N=400\) and \(4000\); these durations decrease inversely with $f$. 

Coherent population activity need not imply strongly super-Poisson single-neuron statistics: synchronized oscillatory spiking networks have been reported with Fano factors close to unity, including \(F_1=0.97\pm0.10\) for pyramidal neurons and \(F_1=1.24\pm0.05\) for interneurons \cite{ardid2010reconciling}. Conversely, homogeneous recurrent integrate-and-fire networks with slow collective fluctuations can exhibit long-window Fano factors of order \(10\) and substantially larger \cite{wieland2015slow}, while bistable or metastable spiking dynamics can generate giant count variability and state durations ranging from hundreds of milliseconds to several seconds \cite{kullmann2022critical,pietras2022mesoscopic}. Strongly synchronous recurrent regimes are also well established \cite{brunel2000dynamics,devalle2017synchrony}, so \(\rho\) need not be small. We therefore regard \(10\,\mathrm{s}\) of post-transient activity as a reasonable default measurement window for stable dynamical regimes and extend the window toward \(100\,\mathrm{s}\) for strongly synchronous, slowly fluctuating, or metastable regimes. 


\section{LIF benchmark parameters}
\label{app:lif_parameters}
All LIF benchmark parameters are given in Table~\ref{tab:lif_parameters}.
\begin{table*}[htbp]
\caption{Typical values and tested ranges for the recurrent LIF benchmark.}
\label{tab:lif_parameters}
\begin{ruledtabular}
\begin{tabular}{llll}
Parameter & Definition & Typical value & Tested range \\
\hline
$S^{EE}$ & E-to-E synaptic strength & 5 & 4--6 \\
$S^{EI}$ & I-to-E synaptic strength & 4.91 & 4--6 \\
$S^{IE}$ & E-to-I synaptic strength & 2 & 1.5--2.5 \\
$S^{II}$ & I-to-I synaptic strength & 4.91 & 4--6 \\
$P^{EE}$ & E-to-E delivery probability & 0.15 & 0.05--0.45 \\
$P^{EI}$ & I-to-E delivery probability & 0.5 & 0.05--0.85 \\
$P^{IE}$ & E-to-I delivery probability & 0.5 & 0.05--0.85 \\
$P^{II}$ & I-to-I delivery probability & 0.4 & 0.05--0.65 \\
$\lambda^{E,\mathrm{ext}}$ & External-to-E rate & 7 kHz & 0.1--10 kHz \\
$\lambda^{I,\mathrm{ext}}$ & External-to-I rate & 7 kHz & Equal to $\lambda^{E,\mathrm{ext}}$ \\
$\tau^{EE}$ & E-to-E synaptic timescale & 4 ms & 1--5 ms \\
$\tau^{EI}$ & I-to-E synaptic timescale & 4.5 ms & 4--5 ms \\
$\tau^{IE}$ & E-to-I synaptic timescale & 1.2 ms & 0.5--2 ms \\
$\tau^{II}$ & I-to-I synaptic timescale & 4.5 ms & 4--5 ms \\
$\tau_{\mathrm{ref}}^E$ & E refractory period & 2 ms & 0.8--2.3 ms \\
$\tau_{\mathrm{ref}}^I$ & I refractory period & 1.6 ms & 0.8--2.3 ms \\
\end{tabular}
\end{ruledtabular}
\end{table*}


\section{EIF robustness figures}
\label{app:eif_results}
\begin{figure}[htbp]
 \centering
 \includegraphics[width=0.99\linewidth]{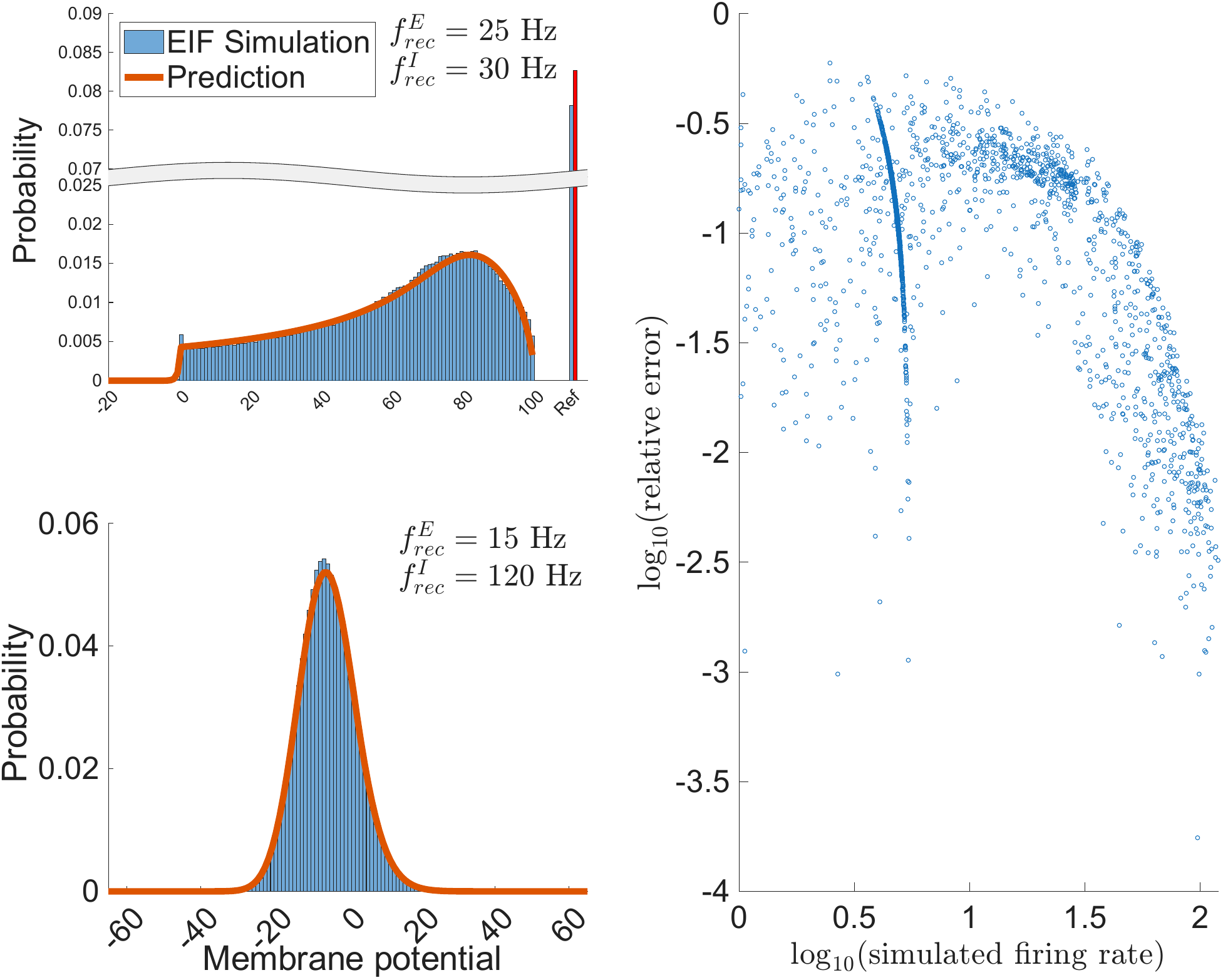}
 \caption{The Type I estimator for a single EIF neuron receiving temporally homogeneous Poisson input. Left: two examples compare the long-time (duration $T=100$~s and step size $\delta t=0.1$~ms) simulated voltage histogram with the Markovian stationary prediction; the narrow bar marked ``Ref'' is the refractory-state probability. Right: relative firing-rate errors for 3,000 parameter configurations plotted against the simulated firing rates on logarithmic axes. Recurrent input rates, refractory periods, and external input rates are varied.}
 \label{fig:eif_single}
\end{figure}

\begin{figure}[htbp]
 \centering
 \includegraphics[width=0.99\linewidth]{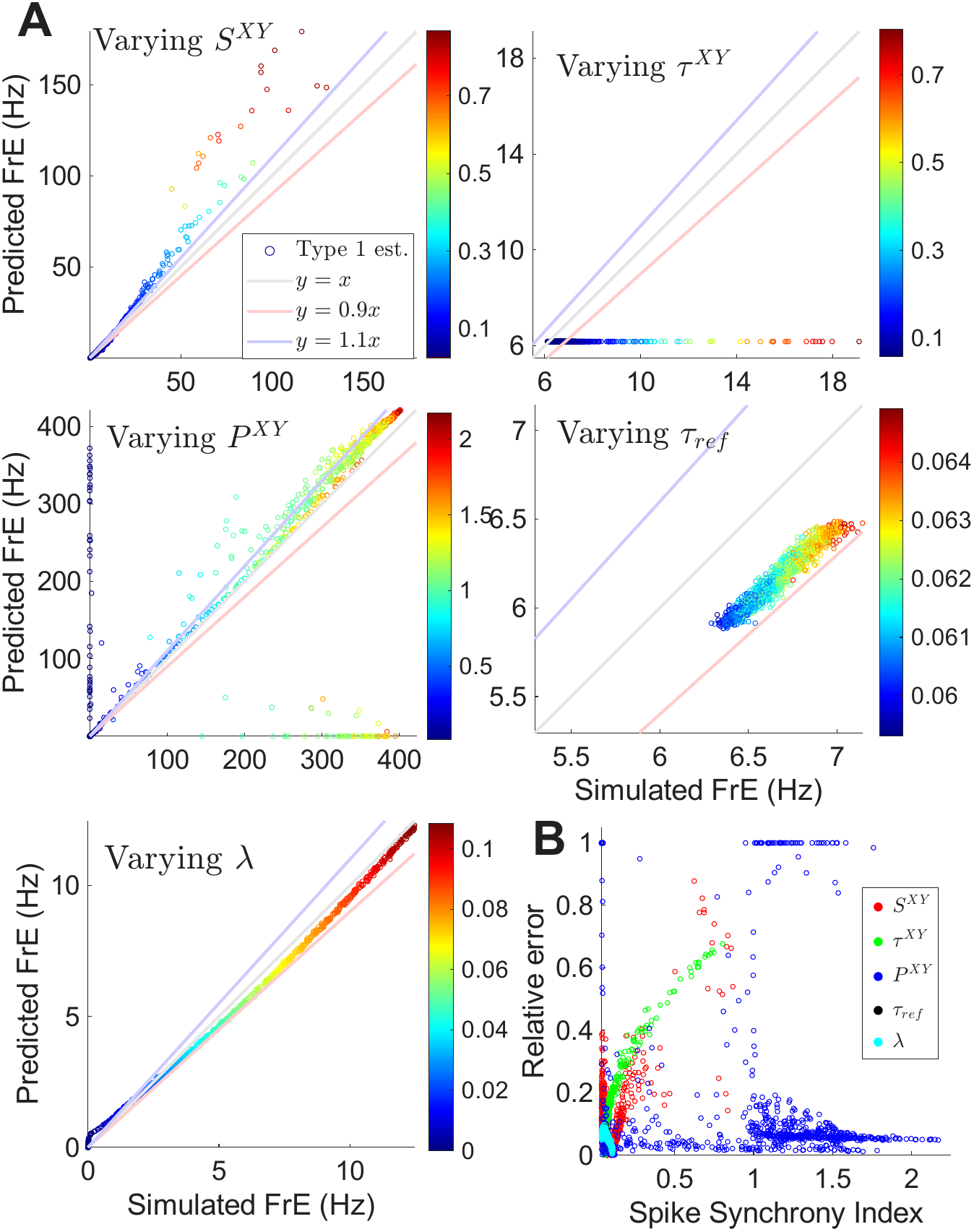}
 \caption{The Type I estimator for recurrent EIF networks with 300 excitatory and 100 inhibitory neurons. (A) Predicted versus simulated excitatory firing rates for 5,000 parameter sets. Parameters have the same meaning as in Fig.~\ref{fig:lif_type1}. (B) Relative error versus SSI, with symbols grouped by the varied parameter family.}
 \label{fig:eif_type1}
\end{figure}

\begin{figure}[htbp]
 \centering
 \includegraphics[width=0.99\linewidth]{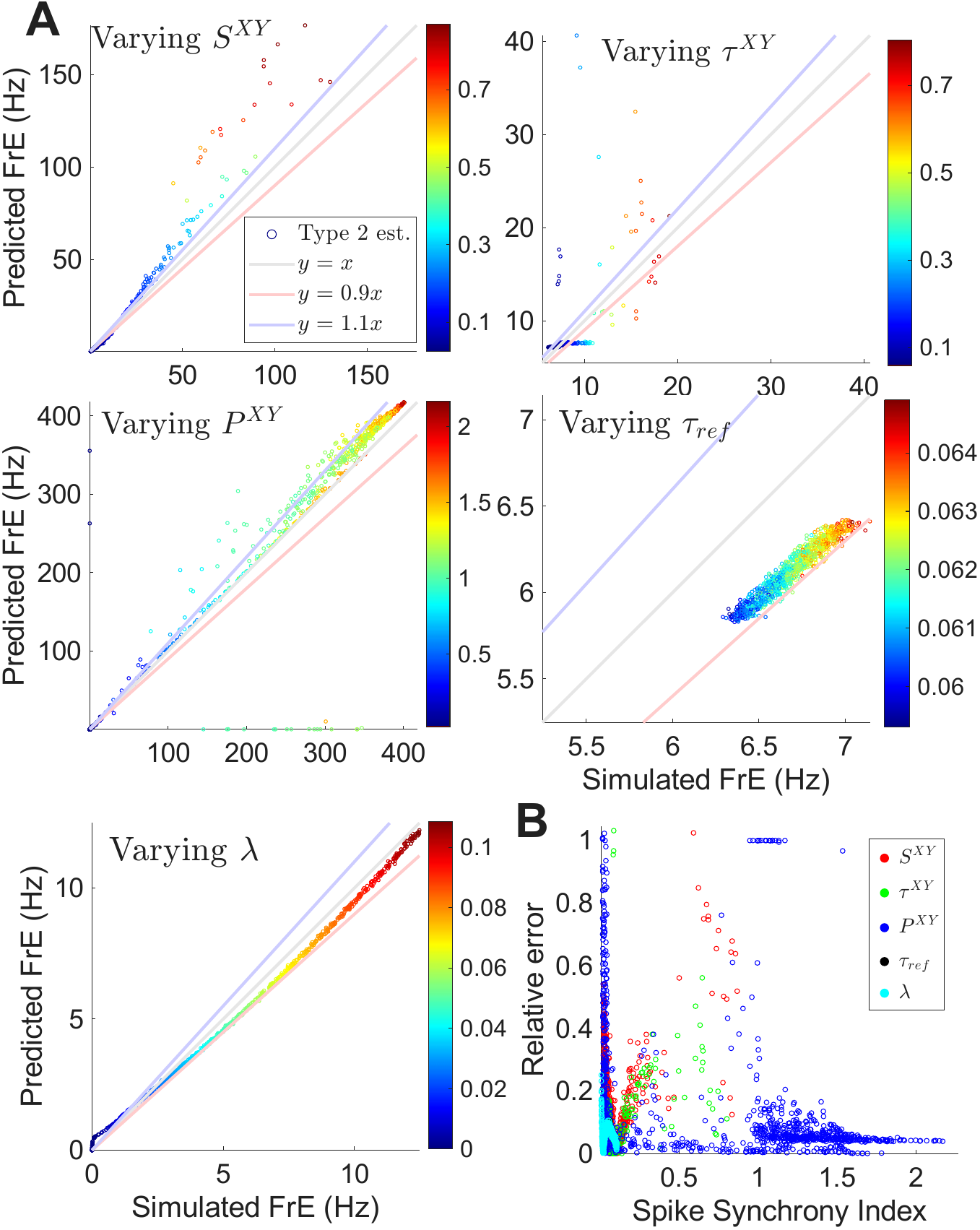}
 \caption{The Type II estimator for the same recurrent EIF networks and 5,000 parameter sets as in Fig.~\ref{fig:eif_type1}. (A) Predicted versus simulated excitatory firing rates for the five parameter families; color encodes SSI. (B) Relative error versus SSI. As in the LIF benchmark, retaining the temporal evolution of the mean synaptic drive improves accuracy in the $\tau^{XY}$ sweep relative to Type I, although outliers remain.}
 \label{fig:eif_type2}
\end{figure}

\begin{figure}[htbp]
 \centering
 \includegraphics[width=0.99\linewidth]{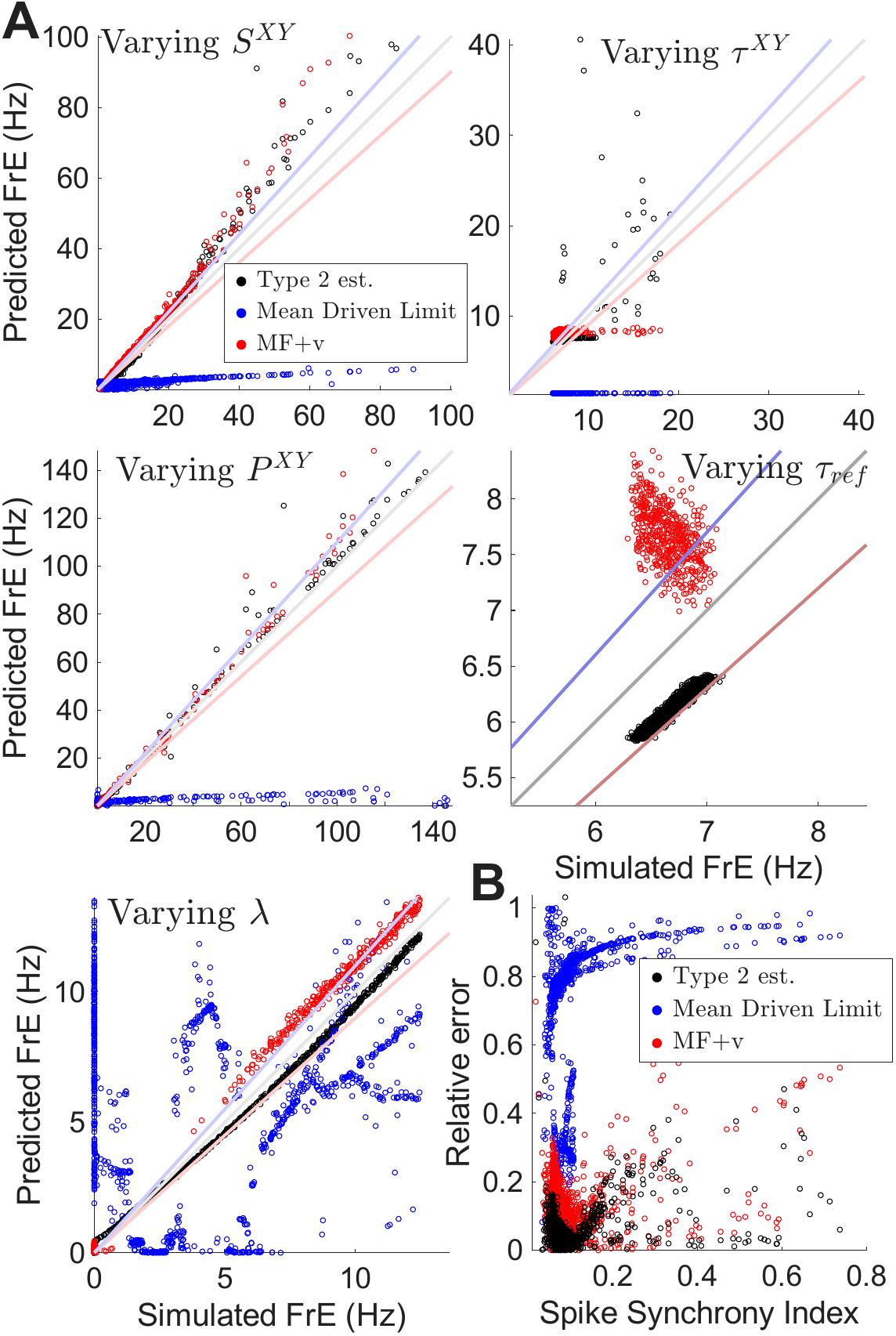}
 \caption{Firing-rate comparison for the Type II estimator, the mean-driven limit, and MF+$v$ in EIF networks with 300 excitatory and 100 inhibitory neurons. (A) Predicted versus simulated excitatory firing rates. The mean-driven limit is not applied to the refractory-period sweep. (B) Relative error versus SSI.}
 \label{fig:eif_compare}
\end{figure}

Results of the Type I and II estimators on EIF networks are shown in Fig.~\ref{fig:eif_single}-\ref{fig:eif_compare}.

\section{Initial-condition-dependent activity branches in LIF networks}
\label{app:branches}

\begin{figure*}[htbp]
 \centering
 \includegraphics[width=0.95\textwidth]{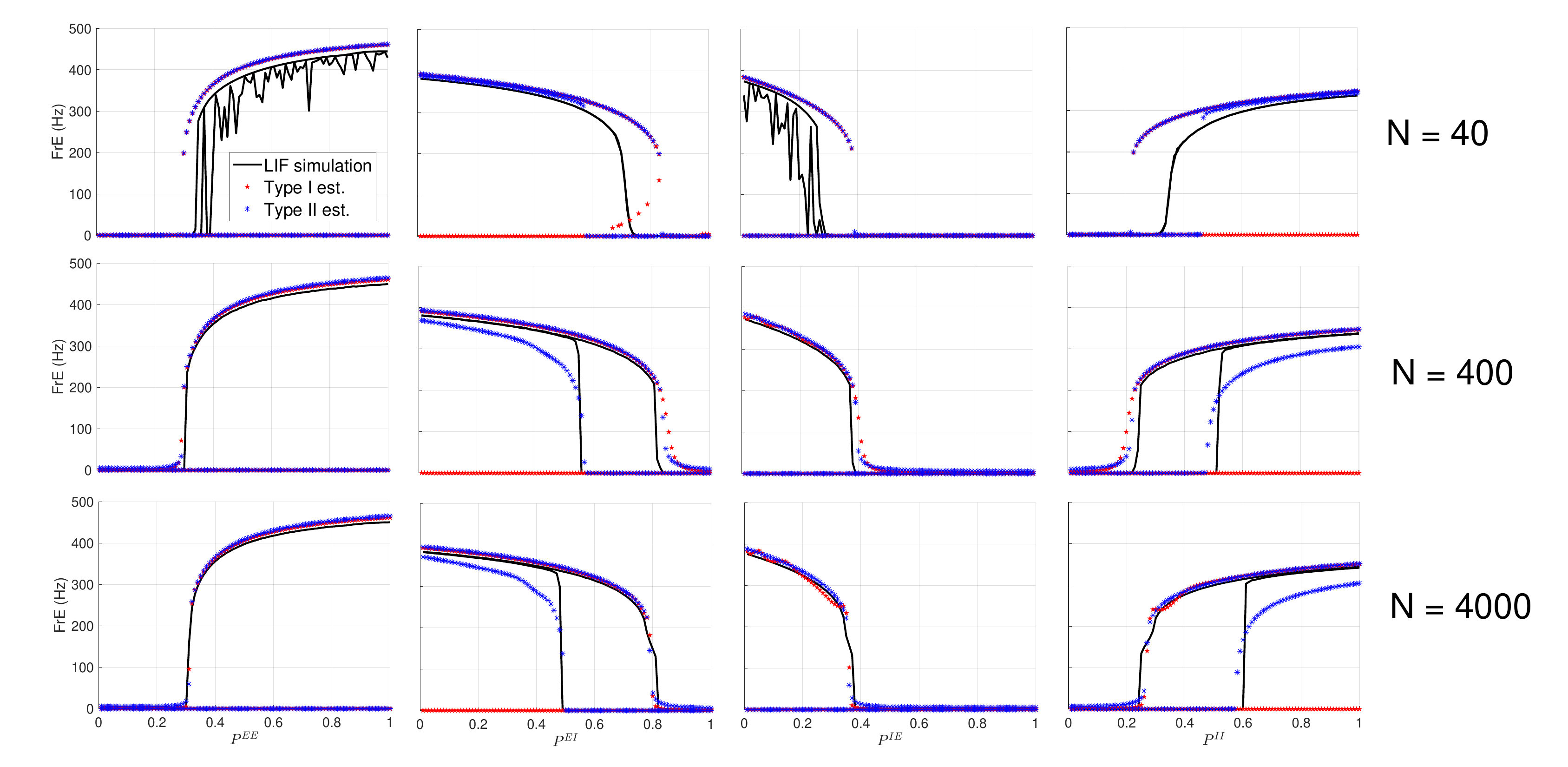}
 \caption{Initial-condition-dependent LIF activity branches at three network sizes. Rows show networks with 30 excitatory and 10 inhibitory neurons ($N=40$), 300 excitatory and 100 inhibitory neurons ($N=400$), and 3,000 excitatory and 1,000 inhibitory neurons ($N=4{,}000$). Columns vary $P^{EE}$, $P^{EI}$, $P^{IE}$, and $P^{II}$ from left to right. Curves and markers compare 10-s direct simulations from different initial conditions with both Type I and Type II estimators. Type II displays both low- and high-activity branches, whereas Type I displays one branch.}
 \label{fig:lif_branches}
\end{figure*}

For the coupling-probability sweeps in Fig.~\ref{fig:lif_branches}, curves and markers compare 10-s direct simulations and reduced calculations initialized from the corresponding low- and high-activity states. Across the displayed sweeps, Type II retains both initial-condition-dependent branches, whereas the Type I iteration converges to only one. This pattern is visible for total network sizes $N\in\{40,400,4{,}000\}$.

Because these results are finite-time parameter sweeps rather than continuation and stability analyses, we refer to initial-condition-dependent activity branches rather than established bifurcations or phase transitions. They may reflect multistability, metastability, or long transients. Type II's closer tracking is consistent with its temporal mean-drive feedback, but the present experiment does not distinguish among these mechanisms.


\clearpage
\bibliography{apssamp}

@article{ardid2010reconciling,
  author  = {Ardid, Salva and Wang, Xiao-Jing and
             Gomez-Cabrero, David and Compte, Albert},
  title   = {Reconciling Coherent Oscillation with Modulation of
             Irregular Spiking Activity in Selective Attention:
             {Gamma}-Range Synchronization between Sensory and
             Executive Cortical Areas},
  journal = {The Journal of Neuroscience},
  year    = {2010},
  volume  = {30},
  number  = {8},
  pages   = {2856--2870},
  doi     = {10.1523/JNEUROSCI.4222-09.2010}
}

@article{wieland2015slow,
  author  = {Wieland, Stefan and Bernardi, Davide and
             Schwalger, Tilo and Lindner, Benjamin},
  title   = {Slow Fluctuations in Recurrent Networks of
             Spiking Neurons},
  journal = {Physical Review E},
  year    = {2015},
  volume  = {92},
  number  = {4},
  pages   = {040901},
  doi     = {10.1103/PhysRevE.92.040901}
}

@article{kullmann2022critical,
  author  = {Kullmann, Richard and Knoll, Gregory and
             Bernardi, Davide and Lindner, Benjamin},
  title   = {Critical Current for Giant {Fano} Factor in Neural
             Models with Bistable Firing Dynamics and
             Implications for Signal Transmission},
  journal = {Physical Review E},
  year    = {2022},
  volume  = {105},
  number  = {1},
  pages   = {014416},
  doi     = {10.1103/PhysRevE.105.014416}
}

@article{pietras2022mesoscopic,
  author  = {Pietras, Bastian and Schmutz, Valentin and
             Schwalger, Tilo},
  title   = {Mesoscopic Description of Hippocampal Replay and
             Metastability in Spiking Neural Networks with
             Short-Term Plasticity},
  journal = {PLOS Computational Biology},
  year    = {2022},
  volume  = {18},
  number  = {12},
  pages   = {e1010809},
  doi     = {10.1371/journal.pcbi.1010809}
}

@article{chang2025minimizing,
  title={Minimizing information loss reduces spiking neuronal networks to differential equations},
  author={Chang, Jie and Li, Zhuoran and Wang, Zhongyi and Tao, Louis and Xiao, Zhuo-Cheng},
  journal={Journal of Computational Physics},
  volume={537},
  pages={114117},
  year={2025},
  publisher={Elsevier},
  doi={10.1016/j.jcp.2025.114117}
}

@article{xiao2021data,
  title={A data-informed mean-field approach to mapping of cortical parameter landscapes},
  author={Xiao, Zhuo-Cheng and Lin, Kevin K and Young, Lai-Sang},
  journal={PLoS Computational Biology},
  volume={17},
  number={12},
  pages={e1009718},
  year={2021},
  publisher={Public Library of Science San Francisco, CA USA},
  doi={10.1371/journal.pcbi.1009718}
}

@article{cai2021model,
  title={Model Reduction Captures Stochastic Gamma Oscillations on Low-Dimensional Manifolds},
  author={Cai, Yuhang and Wu, Tianyi and Tao, Louis and Xiao, Zhuo-Cheng},
  journal={Frontiers in Computational Neuroscience},
  volume={15},
  pages={678688},
  year={2021},
  publisher={Frontiers},
  doi={10.3389/fncom.2021.678688}
}

@article{cai2006kinetic,
  title={Kinetic theory for neuronal network dynamics},
  author={Cai, David and Tao, Louis and Rangan, Aaditya V and McLaughlin, David W},
  journal={Communications in Mathematical Sciences},
  volume={4},
  number={1},
  pages={97--127},
  year={2006},
  publisher={International Press of Boston},
  doi={10.4310/CMS.2006.v4.n1.a4}
}

@article{chariker2016orientation,
  title={Orientation selectivity from very sparse LGN inputs in a comprehensive model of macaque V1 cortex},
  author={Chariker, Logan and Shapley, Robert and Young, Lai-Sang},
  journal={Journal of Neuroscience},
  volume={36},
  number={49},
  pages={12368--12384},
  year={2016},
  publisher={Soc Neuroscience}
}

@article{chariker2022computational,
  title={A computational model of direction selectivity in Macaque V1 cortex based on dynamic differences between ON and OFF pathways},
  author={Chariker, Logan and Shapley, Robert and Hawken, Michael and Young, Lai-Sang},
  journal={Journal of Neuroscience},
  volume={42},
  number={16},
  pages={3365--3380},
  year={2022},
  publisher={Soc Neuroscience}
}

@article{chariker2018rhythm,
  title={Rhythm and synchrony in a cortical network model},
  author={Chariker, Logan and Shapley, Robert and Young, Lai-Sang},
  journal={Journal of Neuroscience},
  volume={38},
  number={40},
  pages={8621--8634},
  year={2018},
  publisher={Soc Neuroscience}
}

@article{li2019well,
  title={How well do reduced models capture the dynamics in models of interacting neurons?},
  author={Li, Yao and Chariker, Logan and Young, Lai-Sang},
  journal={Journal of mathematical biology},
  volume={78},
  number={1--2},
  pages={83--115},
  year={2019},
  publisher={Springer},
  doi={10.1007/s00285-018-1268-0}
}

@article{treves1993meanfield,
    author = {Treves, Alessandro},
    title = {Mean-field analysis of neuronal spike dynamics},
    journal = {Network},
    volume = {4},
    pages = {259--284},
    year = {1993},
    publisher = {IOP Publishing Ltd}
}

@article{wu2023multi,
  title={Multi-band oscillations emerge from a simple spiking network},
  author={Wu, Tianyi and Cai, Yuhang and Zhang, Ruilin and Wang, Zhongyi and Tao, Louis and Xiao, Zhuo-Cheng},
  journal={Chaos: An Interdisciplinary Journal of Nonlinear Science},
  volume={33},
  number={4},
  pages={043121},
  year={2023},
  publisher={AIP Publishing},
  doi={10.1063/5.0106884}
}

@article{zenke2021remarkable,
  title={The remarkable robustness of surrogate gradient learning for instilling complex function in spiking neural networks},
  author={Zenke, Friedemann and Vogels, Tim P},
  journal={Neural computation},
  volume={33},
  number={4},
  pages={899--925},
  year={2021},
  publisher={MIT Press}
}

@article{fitzhugh1961impulses,
  title={Impulses and physiological states in theoretical models of nerve membrane},
  author={FitzHugh, Richard},
  journal={Biophysical journal},
  volume={1},
  number={6},
  pages={445--466},
  year={1961},
  publisher={Elsevier}
}

@article{buice2010systematic,
  title={Systematic fluctuation expansion for neural network activity equations},
  author={Buice, Michael A and Cowan, Jack D and Chow, Carson C},
  journal={Neural computation},
  volume={22},
  number={2},
  pages={377--426},
  year={2010},
  publisher={MIT Press},
  doi={10.1162/neco.2009.02-09-960}
}

@article{schwalger2017towards,
  title={Towards a theory of cortical columns: From spiking neurons to interacting neural populations of finite size},
  author={Schwalger, Tilo and Deger, Moritz and Gerstner, Wulfram},
  journal={PLoS computational biology},
  volume={13},
  number={4},
  pages={e1005507},
  year={2017},
  publisher={Public Library of Science San Francisco, CA USA},
  doi={10.1371/journal.pcbi.1005507}
}

@article{siegle2021survey,
  title={Survey of spiking in the mouse visual system reveals functional hierarchy},
  author={Siegle, Joshua H and Jia, Xiaoxuan and Durand, S{\'e}verine and Gale, Sam and Bennett, Corbett and Graddis, Nile and Heller, Greggory and Ramirez, Tamina K and Choi, Hannah and Luviano, Jennifer A and others},
  journal={Nature},
  volume={592},
  number={7852},
  pages={86--92},
  year={2021},
  publisher={Nature Publishing Group UK London}
}

@article{billeh2020systematic,
  title={Systematic integration of structural and functional data into multi-scale models of mouse primary visual cortex},
  author={Billeh, Yazan N and Cai, Binghuang and Gratiy, Sergey L and Dai, Kael and Iyer, Ramakrishnan and Gouwens, Nathan W and Abbasi-Asl, Reza and Jia, Xiaoxuan and Siegle, Joshua H and Olsen, Shawn R and others},
  journal={Neuron},
  volume={106},
  number={3},
  pages={388--403},
  year={2020},
  publisher={Elsevier}
}

@article{schmidt2018multi,
  title={A multi-scale layer-resolved spiking network model of resting-state dynamics in macaque visual cortical areas},
  author={Schmidt, Maximilian and Bakker, Rembrandt and Shen, Kelly and Bezgin, Gleb and Diesmann, Markus and van Albada, Sacha Jennifer},
  journal={PLOS Computational Biology},
  volume={14},
  number={10},
  pages={e1006359},
  year={2018},
  publisher={Public Library of Science San Francisco, CA USA}
}

@article{markram2015reconstruction,
  title={Reconstruction and simulation of neocortical microcircuitry},
  author={Markram, Henry and Muller, Eilif and Ramaswamy, Srikanth and Reimann, Michael W and Abdellah, Marwan and Sanchez, Carlos Aguado and Ailamaki, Anastasia and Alonso-Nanclares, Lidia and Antille, Nicolas and Arsever, Selim and others},
  journal={Cell},
  volume={163},
  number={2},
  pages={456--492},
  year={2015},
  publisher={Elsevier}
}

@article{bezaire2016interneuronal,
  title={Interneuronal mechanisms of hippocampal theta oscillations in a full-scale model of the rodent CA1 circuit},
  author={Bezaire, Marianne J and Raikov, Ivan and Burk, Kelly and Vyas, Dhrumil and Soltesz, Ivan},
  journal={Elife},
  volume={5},
  pages={e18566},
  year={2016},
  publisher={eLife Sciences Publications, Ltd}
}

@article{izhikevich2008large,
  title={Large-scale model of mammalian thalamocortical systems},
  author={Izhikevich, Eugene M and Edelman, Gerald M},
  journal={Proceedings of the national academy of sciences},
  volume={105},
  number={9},
  pages={3593--3598},
  year={2008},
  publisher={National Acad Sciences}
}

@article{potjans2014cell,
  title={The cell-type specific cortical microcircuit: relating structure and activity in a full-scale spiking network model},
  author={Potjans, Tobias C and Diesmann, Markus},
  journal={Cerebral cortex},
  volume={24},
  number={3},
  pages={785--806},
  year={2014},
  publisher={Oxford University Press}
}

@article{eliasmith2012large,
  title={A large-scale model of the functioning brain},
  author={Eliasmith, Chris and Stewart, Terrence C and Choo, Xuan and Bekolay, Trevor and DeWolf, Travis and Tang, Yichuan and Rasmussen, Daniel},
  journal={science},
  volume={338},
  number={6111},
  pages={1202--1205},
  year={2012},
  publisher={American Association for the Advancement of Science}
}

@article{wang2002probabilistic,
  title={Probabilistic decision making by slow reverberation in cortical circuits},
  author={Wang, Xiao-Jing},
  journal={Neuron},
  volume={36},
  number={5},
  pages={955--968},
  year={2002},
  publisher={Elsevier}
}

@article{brunel1999fast,
  title={Fast global oscillations in networks of integrate-and-fire neurons with low firing rates},
  author={Brunel, Nicolas and Hakim, Vincent},
  journal={Neural computation},
  volume={11},
  number={7},
  pages={1621--1671},
  year={1999},
  publisher={MIT Press}
}

@article{brunel2000dynamics,
  title={Dynamics of sparsely connected networks of excitatory and inhibitory spiking neurons},
  author={Brunel, Nicolas},
  journal={Journal of computational neuroscience},
  volume={8},
  pages={183--208},
  year={2000},
  publisher={Springer}
}

@article{wilson1972excitatory,
  title={Excitatory and inhibitory interactions in localized populations of model neurons},
  author={Wilson, Hugh R and Cowan, Jack D},
  journal={Biophysical journal},
  volume={12},
  number={1},
  pages={1--24},
  year={1972},
  publisher={Elsevier}
}

@article{buice2013dynamic,
  title={Dynamic finite size effects in spiking neural networks},
  author={Buice, Michael A and Chow, Carson C},
  journal={PLoS computational biology},
  volume={9},
  number={1},
  pages={e1002872},
  year={2013},
  publisher={Public Library of Science San Francisco, USA},
  doi={10.1371/journal.pcbi.1002872}
}

@article{hodgkin1952quantitative,
  title={A quantitative description of membrane current and its application to conduction and excitation in nerve},
  author={Hodgkin, Alan L and Huxley, Andrew F},
  journal={The Journal of Physiology},
  volume={117},
  number={4},
  pages={500--544},
  year={1952},
  publisher={Wiley-Blackwell},
  doi={10.1113/jphysiol.1952.sp004764}
}

@article{MontbrioEtAl2015,
  title={Macroscopic description for networks of spiking neurons},
  author={Montbri{\'o}, Ernest and Paz{\'o}, Diego and Roxin, Alex},
  journal={Physical Review X},
  volume={5},
  number={2},
  pages={021028},
  year={2015},
  publisher={APS},
  doi={10.1103/PhysRevX.5.021028}
}

@article{OmurtagEtAl2000,
   author    = {Omurtag, A. and Knight, B.W. and Sirovich, L.},
   title         = {On the simulation of large populations of neurons}, 
   year       = {2000},
   journal   = {J. Comput. Neurosci.},
   volume  = {8},
   pages    = {51--63},
   doi      = {10.1023/A:1008964915724}
}

@article{NykampTranchina2000,
   author    = {Nykamp, D.Q. and Tranchina, D.},
   title         = {A population density approach that facilitates large-scale modeling of neural networks: Analysis and an application to orientation tuning}, 
   year       = {2000},
   journal   = {J. Comput. Neurosci.},
   volume  = {8},
   pages    = {19--50}
}

@Article{pmid16712338,
   Author="Rangan, A. V.  and Cai, D. ",
   Title="{{M}aximum-entropy closures for kinetic theories of neuronal network dynamics}",
   Journal="Phys. Rev. Lett.",
   Year="2006",
   Volume="96",
   Number="17",
   Pages="178101",
   Month="May",
   doi="10.1103/PhysRevLett.96.178101"
}

@book{dayanabbott,
    title = {Theoretical Neuroscience},
    author = {Dayan, Peter and Abbott, L. F.},
    publisher = {MIT Press},
    address = {Cambridge, MA},
    year = {2001},
    isbn = {9780262041997}
}

@article{kass2023,
    author = {Kass, Robert E. and Bong, Heejong and Olarinre, Motolani and Xin, Qi and Urban, Konrad N.},
    title = {Identification of interacting neural populations: methods and statistical considerations},
    journal = {Journal of Neurophysiology},
    volume = {130},
    number = {3},
    pages = {475--496},
    year = {2023},
    doi = {10.1152/jn.00131.2023}
}

@article{mizuseki2013,
    author = {Mizuseki, Kenji and Buzs{\'a}ki, Gy{\"o}rgy},
    title = {Preconfigured, Skewed Distribution of Firing Rates in the Hippocampus and Entorhinal Cortex},
    journal = {Cell Reports},
volume = {4},
number = {5},
pages = {1010--1021},
    year = {2013}
}

@article{baccelli2021pairreplica,
    author = {Baccelli, Fran{\c{c}}ois and Taillefumier, Thibaud},
    title = {The Pair-Replica-Mean-Field Limit for Intensity-based Neural Networks},
    journal = {SIAM Journal on Applied Dynamical Systems},
volume = {20},
number = {1},
pages = {165--207},
    year = {2021},
    doi = {10.1137/20M1331664}
}

@article{kulkani2021benchmark,
    author = {Kulkarni, Shruti R and Parsa, Maryam and Mitchell, J Parker and Schuman, Catherine D},
    title = {Benchmarking the performance of neuromorphic and spiking neural network simulators},
    journal = {Neurocomputing},
volume = {447},
pages = {145--160},
    year = {2021},
    doi = {10.1016/j.neucom.2021.03.028}
}

@article{brian2genn,
    author = {Stimberg, Marcel and Goodman, Dan F M and Nowotny, Thomas},
    title = {Brian2GeNN: accelerating spiking neural network simulations with graphics hardware},
    journal = {Scientific Reports},
volume = {10},
pages = {410},
    year = {2020},
    doi = {10.1038/s41598-019-54957-7}
}

@article{schmitt2023simulation,
    author = {Schmitt, Felix Johannes and Rostami, Vahid and Nawrot, Martin Paul},
    title = {Efficient parameter calibration and real-time simulation of large-scale spiking neural networks with GeNN and NEST},
    journal = {Frontiers in Neuroinformatics},
volume = {17},
pages = {941696},
    year = {2023},
    doi = {10.3389/fninf.2023.941696}
}

@article{baladron2012meanfield,
  author = {Baladron, Javier and Fasoli, Diego and Faugeras, Olivier and Touboul, Jonathan},
  title = {Mean-field description and propagation of chaos in networks of {Hodgkin--Huxley} and {FitzHugh--Nagumo} neurons},
  journal = {The Journal of Mathematical Neuroscience},
  volume = {2},
  number = {1},
  pages = {10},
  year = {2012},
  doi = {10.1186/2190-8567-2-10}
}

@article{nicola2013twodim,
  author = {Nicola, Wilten and Campbell, Sue Ann},
  title = {Mean-field models for heterogeneous networks of two-dimensional integrate and fire neurons},
  journal = {Frontiers in Computational Neuroscience},
  volume = {7},
  pages = {184},
  year = {2013},
  doi = {10.3389/fncom.2013.00184}
}

@article{augustin2017lowdim,
  author = {Augustin, Moritz and Ladenbauer, Josef and Baumann, Fabian and Obermayer, Klaus},
  title = {Low-dimensional spike rate models derived from networks of adaptive integrate-and-fire neurons: Comparison and implementation},
  journal = {PLoS Computational Biology},
  volume = {13},
  number = {6},
  pages = {e1005545},
  year = {2017},
  doi = {10.1371/journal.pcbi.1005545}
}

@article{zerlaut2018mesoscopic,
  author = {Zerlaut, Yann and Chemla, Sandrine and Chavane, Fr{\'e}d{\'e}ric and Destexhe, Alain},
  title = {Modeling mesoscopic cortical dynamics using a mean-field model of conductance-based networks of adaptive exponential integrate-and-fire neurons},
  journal = {Journal of Computational Neuroscience},
  volume = {44},
  number = {1},
  pages = {45--61},
  year = {2018},
  doi = {10.1007/s10827-017-0668-2}
}

@article{carlu2020complex,
  author = {Carlu, Michele and Chehab, Olivier and Dalla Porta, Leonardo and Depannemaecker, Damien and H{\'e}ric{\'e}, Costantino and Jedynak, Maciej and K{\"o}ksal Ers{\"o}z, Eylem and Muratore, Paolo and Souihel, Sarah and Capone, Cristiano and Zerlaut, Yann and Destexhe, Alain and di Volo, Matteo},
  title = {A mean-field approach to the dynamics of networks of complex neurons, from nonlinear integrate-and-fire to {Hodgkin--Huxley} models},
  journal = {Journal of Neurophysiology},
  volume = {123},
  number = {3},
  pages = {1042--1051},
  year = {2020},
  doi = {10.1152/jn.00399.2019}
}

@article{osborne2022ndimensional,
  author = {Osborne, Hugh and de Kamps, Marc},
  title = {A numerical population density technique for {N}-dimensional neuron models},
  journal = {Frontiers in Neuroinformatics},
  volume = {16},
  pages = {883796},
  year = {2022},
  doi = {10.3389/fninf.2022.883796}
}

@article{richardson2010shotnoise,
  author = {Richardson, Magnus J. E. and Swarbrick, Rupert},
  title = {Firing-rate response of a neuron receiving excitatory and inhibitory synaptic shot noise},
  journal = {Physical Review Letters},
  volume = {105},
  number = {17},
  pages = {178102},
  year = {2010},
  doi = {10.1103/PhysRevLett.105.178102}
}

@article{devalle2017synchrony,
  author = {Devalle, Federico and Roxin, Alex and Montbri{\'o}, Ernest},
  title = {Firing rate equations require a spike synchrony mechanism to correctly describe fast oscillations in inhibitory networks},
  journal = {PLoS Computational Biology},
  volume = {13},
  number = {12},
  pages = {e1005881},
  year = {2017},
  doi = {10.1371/journal.pcbi.1005881}
}

@article{ocker2017nonlinear,
  author = {Ocker, Gabriel Koch and Josi{\'c}, Kre{\v{s}}imir and Shea-Brown, Eric and Buice, Michael A.},
  title = {Linking structure and activity in nonlinear spiking networks},
  journal = {PLoS Computational Biology},
  volume = {13},
  number = {6},
  pages = {e1005583},
  year = {2017},
  doi = {10.1371/journal.pcbi.1005583}
}

@article{rangan2008conductance,
  author = {Rangan, Aaditya V. and Kova{\v{c}}i{\v{c}}, Gregor and Cai, David},
  title = {Kinetic theory for neuronal networks with fast and slow excitatory conductances driven by the same spike train},
  journal = {Physical Review E},
  volume = {77},
  number = {4},
  pages = {041915},
  year = {2008},
  doi = {10.1103/PhysRevE.77.041915}
}

@article{meunier1992dimreductionHH,
    author = {Meunier, Claude},
    title = {Two and three dimensional reductions of the Hodgkin-Huxley system: separation of time scales and bifurcation schemes},
    journal = {Biological Cybernetics},
    volume = {67},
    pages = {461--468},
    year = {1992},
    month = "Sep",
    doi = {10.1007/BF00200990}
}

@article{rinzel1985excitation,
    author = {Rinzel, John},
    title = {Excitation dynamics: insights from simplified membrane models},
    journal = {Federal Proceedings},
    volume = {44},
    number = {15},
    pages = {2944-2946},
    year = {1985},
    month = "Dec"
}

@ARTICLE{Fourcaud-Trocme2003-lm,
  title     = "How spike generation mechanisms determine the neuronal response
               to fluctuating inputs",
  author    = "Fourcaud-Trocm{\'e}, Nicolas and Hansel, David and van
               Vreeswijk, Carl and Brunel, Nicolas",
  journal   = "J. Neurosci.",
  publisher = "Society for Neuroscience",
  volume    =  23,
  number    =  37,
  pages     = "11628--11640",
  month     =  dec,
  year      =  2003
}

\end{document}